\documentclass[conference]{IEEEtran}

\usepackage{algorithmic}
\usepackage{graphicx}
\usepackage{textcomp}
\usepackage{xcolor}
\let\labelindent\relax
\usepackage{enumitem}
\usepackage{booktabs}
\usepackage{adjustbox}
\usepackage{float}
\usepackage{listings}
\usepackage{subcaption}
\usepackage{amsmath,amsfonts}
\usepackage{url}
\usepackage{hyperref}
\usepackage{comment}

\usepackage[capitalise]{cleveref}
\AtBeginEnvironment{appendices}{\crefalias{section}{appendix}}

\newcommand{\sys}{TorMult} 
\newcommand{\cosys}{C-TorMult} 
\newcommand{\desys}{D-TorMult} 
\begin{document}

\date{}

\title{\Large \bf TorMult: Introducing a Novel Tor Bandwidth Inflation Attack}


{\author
	{\IEEEauthorblockN{Christoph Sendner\IEEEauthorrefmark{1},
			Jasper Stang\IEEEauthorrefmark{1},
			Alexandra Dmitrienko\IEEEauthorrefmark{1},
			Raveen Wijewickrama\IEEEauthorrefmark{2}, and
			Murtuza Jadliwala\IEEEauthorrefmark{2}}
		\IEEEauthorblockA{\IEEEauthorrefmark{1}University of W\"urzburg, Germany\\
		}
		\IEEEauthorblockA{\IEEEauthorrefmark{2}University of Texas at San Antonio, USA}
}}

\maketitle

\begin{abstract}
The Tor network is the most prominent system for providing anonymous communication to web users, with a daily user base of 2 million users. 
However, since its inception, it has been constantly targeted by various traffic fingerprinting and correlation attacks aiming at deanonymizing its users. 
A critical requirement for these attacks is to attract as much user traffic to adversarial relays as possible, which is typically accomplished by means of bandwidth inflation attacks. 
This paper proposes a new inflation attack vector in Tor, referred to as \sys{}, which enables inflation of measured bandwidth. The underlying attack technique exploits resource sharing among Tor relay nodes and employs a cluster of attacker-controlled relays with coordinated resource allocation within the cluster to deceive bandwidth measurers into believing that each relay node in the cluster possesses ample resources. 
We propose two attack variants,~\cosys{} and \desys{}, and test both versions in a private Tor test network. 
Our evaluation demonstrates that an attacker can inflate the measured bandwidth by a factor close to $n$ using \cosys{} and nearly half $n*N$ using \desys{}, where $n$ is the size of the cluster hosted on one server and $N$ is the number of servers. Furthermore, our theoretical analysis reveals that gaining control over half of the Tor network's traffic can be achieved by employing just 10 dedicated servers with a cluster size of 109 relays running the \sys{} attack, each with a bandwidth of 100MB/s. 
The problem is further exacerbated by the fact that Tor not only allows resource sharing but, according to recent reports, even promotes it. 
\end{abstract}

\section{Introduction}
\label{sec:introduction}
In an era where Internet-based service providers, advertisers, and government agencies are increasingly focused on tracking web users and monitoring their activities, privacy-enhancing technologies have emerged as a powerful defense against malicious web-based user tracking. Among these technologies, anonymous communication networks, such as \emph{Tor}~\cite{torProject} (short for \emph{The Onion Router}), have demonstrated their effectiveness and are rapidly gaining popularity.
Tor relies on more than 7000 volunteer-run servers, called onion \emph{routers} or \emph{relays}, to route encrypted web traffic for nearly two million users daily through anonymity-preserving circuits comprised of anywhere between 3 to 8 (normally, 3) of these relays.

Due to its open-source nature, transparent development process, and appeal to privacy-concerned web users, Tor has attracted considerable attention in the research community. Researchers have extensively investigated various attacks, including deanonymization attempts, aiming to gain insights into the network traffic flowing through Tor. 
Some of the significant attack vectors on Tor include website fingerprinting \cite{cai2012touching,hayes2016k,li2018measuring,panchenko2016website,sirinam2018deep,wang2013improved}, routing~\cite{wails2018tempest,tan2016data,sun2015raptor}, end-to-end correlation~\cite{johnson2013users,ling2013protocol,nasr2018deepcorr},
congestion~\cite{evans2009practical,geddes2013low,jansen2014sniper}, and side channel \cite{hopper2010much,mittal2011stealthy} attacks. These attacks pose a realistic threat to Tor users' security and privacy, as there have been multiple reports of their usage in the wild by state-sponsored entities  \cite{the_guardian_2013,tor_blog_2015,tor_blog_2014}.
However, one common requirement for all these attacks is that the attacker needs to attract as much user traffic as possible to its servers or relays, all while using as few resources as possible. How an attacker can effectively accomplish this pre-requisite is our main focus in this paper.

As an integral part of its regular operation, the Tor network maintains a list of active relays in the form of a \emph{consensus file} containing useful self-reported and measured state information about the relays in the network, including the estimated bandwidth available at each relay. Tor clients employ a \emph{path selection algorithm} which utilizes the most recent consensus file as input. This algorithm forms communication circuits composed of relays through which user data is transmitted. To determine the available bandwidth at the relays, Tor employs its \emph{bandwidth-scanning} techniques that involve the establishment of a two-hop measurement circuit through the relay and performing data transfer (downloading/uploading) to a web server along this path. By utilizing this approach, Tor can measure the bandwidth at the relays and incorporate this information into the consensus file.

The issue is that measurement circuits created and used during bandwidth measurements are built using only two relays, while traffic needs to pass through at least three relays to ensure anonymity \cite{Why3Hops}. This makes the traffic generated by bandwidth measurers easily distinguishable from the regular user traffic. 
Once such measurement circuits are detected, attackers can stop serving user traffic and utilize their entire available bandwidth to forward only measurement traffic, thereby inflating their available bandwidth during measurement. It has been shown in the literature that by employing such a strategy, attackers were able to increase their measured bandwidth and probability of being selected for forwarding users' traffic up to 177× \cite{thill2014hidden,johnson2017peerflow}. 
Another straightforward strategy for an inflation attack involves malicious relays misreporting their available bandwidth in their descriptors \cite{bauer2007low}. These relays attempt to increase their share of users' traffic by providing false and exaggerated reports, regardless of their actual bandwidth capabilities. 

Our study of Tor's bandwidth measurement mechanisms unveils another critical weakness. Currently, nothing prevents relay operators from deploying them on shared resources, such as physical machines or network links. For instance, several attacker-controlled relay nodes can be hosted using virtual machines executed on a single physical host, or, even if residing on separate physical machines, they can share a network link. We find out that this fact can be leveraged by attackers to even further inflate their measured bandwidth by devoting the entire resources of a physical host (with co-resident relays) and/or of the network link to the currently detected measurement traffic while dropping all the user traffic targeting not just one, but \emph{all} attacker-controlled relay nodes, thus overall increasing the impact of previously known inflation attacks (with the factor of up to 177× \cite{thill2014hidden,johnson2017peerflow})  by an additional factor equal to the count of attacker-controlled relays in the cluster. 
The effectiveness of our attack diminishes when bandwidth scanners simultaneously measure two or more malicious relays within the cluster. However, the impact of such co-measurements remains limited due to the small number of bandwidth scanners responsible for measuring the entire Tor network. Consequently, the likelihood of such co-measurements occurring is low. Yet, it is interesting to explore how the number of bandwidth scanners influences the inflation factor of the proposed attack.

\noindent\textbf{Contributions.} Building upon the insights gained from the above-mentioned observations, this paper provides the following contributions:
\begin{itemize}[leftmargin=*]
    \item \textbf{Novel inflation attack strategy:}  We introduce and investigate \emph{\sys{}}, a novel inflation attack strategy for the Tor network. The core idea is to leverage resource sharing within a cluster of attacker-controlled relays, allowing for the dynamic allocation of networking and computational resources of the \emph{entire cluster} to a relay currently being measured. Our inflation strategy can be combined with other known strategies (e.g., false and exaggerated bandwidth self-reporting~\cite{bauer2007low}, or dropping user traffic during bandwidth measurements~\cite{thill2014hidden,johnson2017peerflow}), thus providing an additional inflation factor and further reducing resources required for attracting a sufficiently large amount of user traffic. 
    \item \textbf{Two \sys{} attack variations:} We propose two attack variants: (i) \emph{\cosys}, and (ii) \emph{\desys}. In~\cosys, an attacker deploys \textbf{C}o-resident relays on a single physical host and dynamically assigns the entire host's computational and networking resources to the measured relay. 
   In \desys{}, attacker-controlled relays share the network link with a \textbf{D}edicated server, to which all the measurement traffic, once detected, is diverted. As a result, each malicious relay (measured by the bandwidth scanner) can claim the total bandwidth available on the shared network link. 

   \item \textbf{Attack Evaluation:} We evaluated both attack variations through tests conducted in a small private Tor test network\footnote{Attack tests on real Tor network were disapproved by the Tor research safety board} based on Chutney \cite{chutney}, and are able to empirically demonstrate an inflation factor very close to $n$ and $n*N$ for~\cosys ~and~\desys, respectively. 
   To study the effect of diminishing inflation due to co-measurements likely to happen in larger networks, we studied the probability of such co-measurements based on data collected from the real Tor network (bandwidth files) and performed theoretical modeling. 
   Our analysis reveals that the inflation factor grows almost linearly to $n$ with up to 15 relays. 
   However, the attack's potential diminishes to at least 76\% of the linear growth as the number of relays in a cluster increases.
    
    \item \textbf{Studying the resilience of existing bandwidth scanning alternatives:} We investigate the resilience of other bandwidth measurement techniques proposed in the literature against \sys{} and other inflation attack techniques and identify pitfalls and promising directions. 
    For instance, we show (experimentally) that the state-of-the-art Machine Learning (ML)-based method \cite{SAMNNModel} can effectively (with F1 score 99\%) and quickly (in less than 0.5 $ms$) distinguish user and measurement traffic even in a 3-hop measurement circuit by analyzing just headers and metrics such as inter-arrival time of transmitted packets. This proves that increasing the number of relays in a measurement circuit isn't a viable measurement alternative. From our analysis, it follows that only methods that eliminate measurement traffic altogether but rather leverage user traffic for bandwidth estimations, such as \cite{snader2009eigenspeed}, have the potential to provide an inflation-resilient bandwidth measurement solution. Yet, existing solutions of this kind have limitations (e.g., vulnerable to Sybil attacks) that need to be addressed in future work. We also devise less intrusive mitigation strategies, such as co-residency and measurement anomaly detection, that can be used with currently deployed bandwidth measurement methods. 

    \item \textbf{Analysis of Tor:} 
   In order to investigate the potential usage of \sys{} in the Tor network by real attackers, we analyzed the Tor network. Our analysis revealed specific characteristics displayed by certain Tor relays that could potentially serve as indicators of an ongoing \sys{} attack. For instance, we discovered that some relay families offer similar bandwidth and share the same uptime, while others leverage alternate port numbers, which might imply the co-resident execution of Tor relay clusters. 
  While we lack the means to definitively confirm the malicious nature of those relays or relay families, we propose a strategy that bandwidth scanning services could employ to validate our hypothesis.
\end{itemize}

In summary, the paper contributes a novel inflation attack strategy, evaluates its effectiveness, explores the resilience of existing bandwidth scanning techniques, and provides an analysis of the Tor network to identify potential indicators of ongoing attacks. Overall, our work strongly advocates for the pursuit of additional research focused on the development of inflation-resilient bandwidth measurement techniques. 

\textbf{Outline.} The remaining part of the paper is organized as follows. \Cref{sec:background} provides a concise overview of background information on Tor. In~\Cref{sec:attack}, we define our adversary model and introduce underlying attack techniques. The evaluation results of our attack are presented in~\Cref{sec:evaluation}, accompanied by additional insights gained on Tor in~\Cref{sec:insights}. \Cref{sec:res_alt_solutions} discusses the resilience of our attack in relation to existing solutions and explores potential countermeasures. Furthermore, \Cref{sec:related_works} delves into related works focusing on Tor inflation attacks. Finally, we conclude our findings in \Cref{sec:conclusion}. 
\section{Background}
\label{sec:background}
In this section, we introduce the technical background related to Tor and its bandwidth measurement protocols.

\noindent\textbf{Tor Relays and Circuits.} Tor users, also known as \emph{clients}, select three or more relays randomly from the Tor network to establish a communication circuit. The initial hop of this circuit, also known as the \emph{guard relay}, serves as the entry point for client connections and typically remains in use for 2-3 months before being rotated. To be considered a guard relay, it meets specific criteria, including stability, sufficient uptime (typically 8-68 days \cite{TorLifeCycle2013}), and a minimum bandwidth of 2 \emph{Mbps}.
The concluding relay in this anonymity-preserving communication circuit is known as the \emph{exit relay}. It represents the last hop, responsible for transmitting the client's traffic to its desired destination on the Internet, beyond the Tor network. Consequently, the IP address visible to service providers is that of the exit relay, ensuring the client's IP address remains concealed.
The relays between the guard and the exit relays, known as \emph{middle} relays, serve as intermediate hops within the circuit. They simply forward the traffic in both directions, enabling communication between a Tor client and its intended destination on the Internet.   
Tor relays are run by a global community of volunteers known as \emph{relay operators}. One such operator can provide multiple relays grouped into \emph{families}. The Tor network also allows for the hosting of two Tor relays using the same public IP address \cite{relayops}. 

\noindent\textbf{Directory and Bandwidth Authorities.} When establishing connections, Tor clients select relays for a circuit from a list supplied by specialized entities known as \emph{Directory Authorities (DAs)}. These DAs maintain an up-to-date roster of active relays, including their flag status (guard, middle, or exit) and bandwidth information. The DAs update this list every hour, and the consensus among the DAs is cryptographically signed and made available for clients.

A subset of these DAs, typically six in number, also serve as \emph{Bandwidth Authorities (BAs)}. These BAs are tasked with the additional responsibility of conducting active measurements to estimate the available bandwidth of Tor relays. 
It is worth noting that these six BAs are responsible for measuring more than 7000 relays (with each BA typically running 4-5 parallel measurement threads). The BAs have a limited timeframe of approximately 60 minutes to complete the measurement round and to include results in the subsequent consensus file \cite{tor-consensus}. 
Furthermore, there are occasions where BAs may be temporarily out of service for several hours or even up to a few months due to various reasons such as updates, deployment issues, or ongoing attacks. Consequently, many measurements may be incomplete or missing during these periods.  

\noindent\textbf{Bandwidth Measurements in Tor.} 
The initial bandwidth scanning mechanism used by BAs was the \emph{TorFlow} scanner in 2011 \cite{perry2009torflow,adrienle_2011}. However, it has been gradually replaced by the \emph{Simple Bandwidth Scanner (SBWS)}~\cite{SBWS} mechanism since 2018~\cite{bandwidthTimeline}. Our focus in this work primarily revolves around SBWS as it has completely replaced TorFlow by now~\cite{bandwidthTimeline}.

Both TorFlow and SBWS methods employ bandwidth measurement by downloading large files multiple times from a web server through two-hop circuits. It is important to note that circuits with two relays do not provide anonymity~\cite{Why3Hops}, hence, the measurement traffic can be trivially distinguished from user traffic. 
In SBWS, the first relay in the circuit is the target relay being measured, while the second relay is a randomly selected exit relay that is at least twice as fast as the target relay \cite{darir2022mleflow}. The web-server provides a 1GB file over HTTPS for the SBWS scanner to download. The SBWS scanner randomly selects a byte range between 0 and 1GB, typically in 16MiB increments, such as 0-16MiB or 16MiB-32MiB, and so on. If the download completes in less than 5 seconds, the range is expanded, and if it takes more than 10 seconds, the range is reduced. Once an appropriate range is determined, five files of appropriate size, meaning they can be downloaded within the time range of 5-10 seconds, are downloaded and the number of bytes and the time taken for each file download is recorded \cite{sbws_how_works}. As such, it follows that one measurement takes at least 25 seconds. 

It is noteworthy that the bandwidth files generated by both SBWS and TorFlow methods contain only a single timestamp per measurement, indicating the end of the measurement process. However, they do not include the starting timestamp or the duration of the measurement. This omission is significant because it prevents the deterministic reproduction of the measurement timeline from the bandwidth files. This information would be valuable, for instance, when identifying co-measured relays (as discussed in \cref{sub:inflation_factor}). Once valid measurements are obtained, both SBWS and TorFlow incorporate them into the bandwidth file, which is subsequently forwarded to the DAs for aggregation.

\noindent\textbf{Building Consensus.} Bandwidth measurement files, also known as BA's votes, are processed by DAs to aggregate the results and compute the bandwidth weights in accordance with the Tor directory protocol~\cite{DA-protocol}. 
In essence, the consensus weight for each relay is calculated based on its self-reported bandwidth and the bandwidth measured by BAs using bandwidth scanners. It is important to note that SBWS computes bandwidth weights relative to all Tor relays. The resulting consensus file is signed by all the DAs and then distributed to Tor clients.

Despite the presence of BAs that assist in cross-verifying self-reported relay bandwidth values by conducting actual measurements, an attacker can still effectively execute inflation attacks through a selective denial-of-service (DoS) strategy. In this scenario, a malicious relay only responds to measurement traffic while dropping all other user traffic \cite{johnson2017}. These attacks are feasible due to the publicly available identity information of the BAs, such as their IPv4 and IPv6 addresses \cite{tor_DAs}. Consequently, the detection of measurement traffic originating from the BAs becomes straightforward upon the arrival of the first packet.

\section{\sys~ Attacks} 
\label{sec:attack}

In this section, we start by introducing the adversary model and discussing the capabilities of the attacker. Following that, we present two variants of \sys{}, namely the co-resident relay attack (\cosys) and the dedicated server attack (\desys).

\subsection{Adversary Model \& Capabilities}
\label{sub:advmodel}
\vspace{-0.20cm}
We consider an adversary who controls multiple instances of Tor relays, either as physical machines, virtual machines, or processes/dockers in the same machine. The adversary controls a machine with multiple public IP addresses. For the \desys~variation, the adversary additionally controls the network router or gateway between the Internet and the servers running the relays.

Furthermore, we assume that the attacker is able to distinguish measurement and user traffic in real-time. 
Currently, this can be trivially accomplished in Tor, as SBWS scanners utilize 2-hop circuits for the measurements (\cref{sec:background}). In \cref{traffic_classification}, we also show that Machine Learning techniques can effectively distinguish user and measurement traffic in 3-hop circuits. 

We further assume that the adversary possesses the capability to promptly forward each type of traffic to a preferred Tor relay instance with minimal delay. Specifically, the adversary directs measurement traffic to a high-resource, high-bandwidth relay while diverting other traffic to a low-resource relay or dropping it altogether with the goal of achieving bandwidth inflation. 

\subsection{High-level idea and attack variations}
\label{sub:high_level_attack}
\vspace{-0.20cm}

In this study, we present and explore \sys{}, a novel tactic for carrying out inflation attacks on the Tor network. Our main concept revolves around utilizing resource sharing among a group of relays under the control of the attacker. This enables the cluster to dynamically allocate networking and computational resources to a relay whose bandwidth is currently being assessed. By combining our inflation strategy with other established techniques, such as false and exaggerated bandwidth self-reporting or selectively dropping user traffic during bandwidth measurements, we can amplify the inflation effect and minimize the resources needed to attract a substantial volume of user traffic.

We introduce two versions of~\sys{}, namely~\cosys{} and~\desys{}.~\cosys{} capitalizes on the utilization of shared bandwidth and computational resources from a single server among all the relays hosted on it. In contrast,~\desys{} takes this concept to the next level by expanding the distribution of resources across a network that operates behind a router.

\subsection{\cosys~- \sys{} with Co-Residing Relays}
\label{sub:attack1_design}

\begin{figure}[t]
\vspace*{-0.3cm}
\centering
\includegraphics[width=0.85\linewidth]{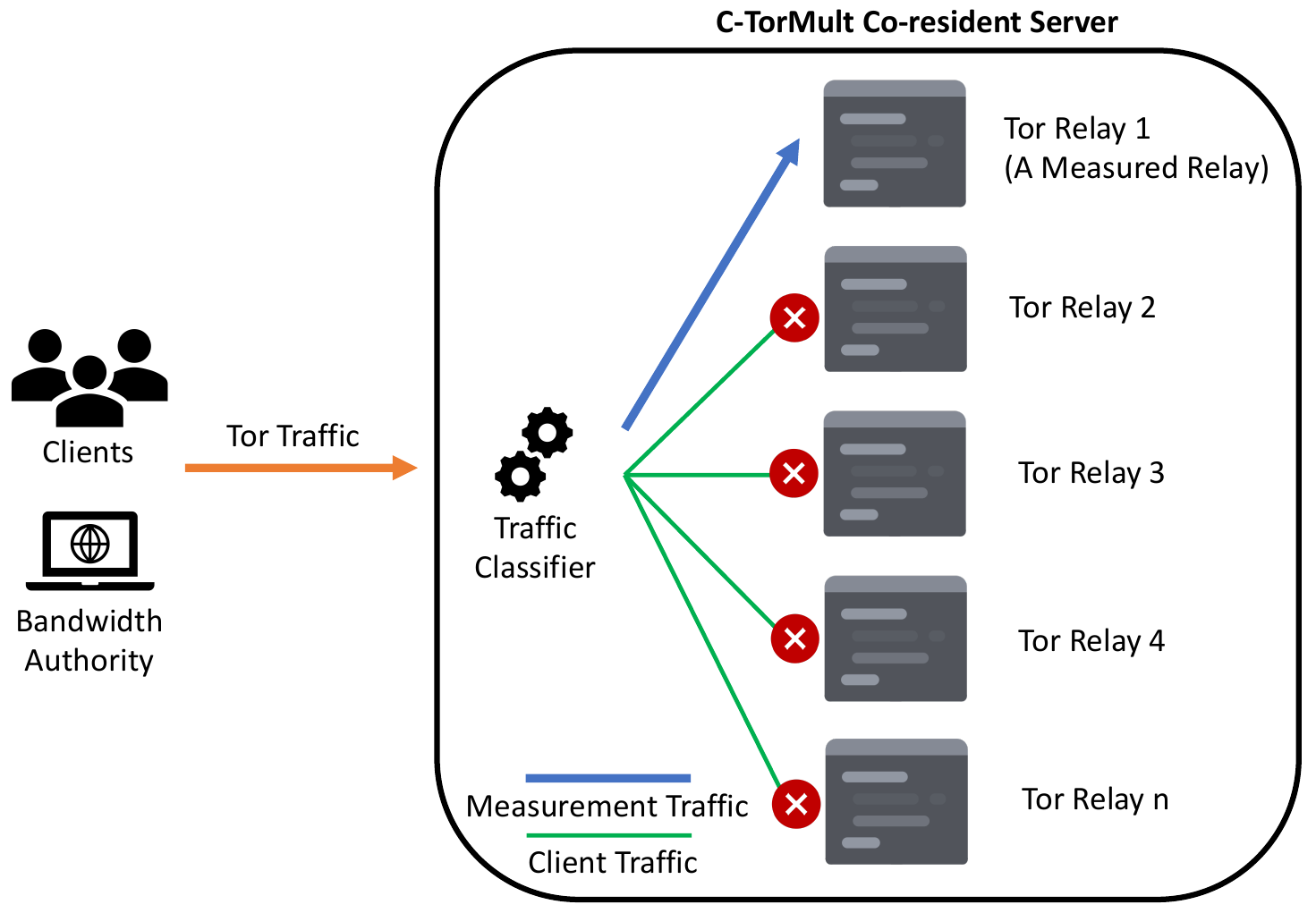}
\vspace*{-0.2cm}
\caption{\cosys: \sys{} with Co-Resident Relays.}
\label{fig:attack_1}
\vspace*{-0.375cm}
\end{figure}

\vspace{-0.2cm}
\noindent\textbf{Design.} As shown in \cref{fig:attack_1}, the attacker operates several \textbf{\emph{co-residing}} Tor relays as a cluster within a machine (i.e., server) in the~\cosys{} attack.
Each co-resident relay can be operated as an additional CPU process, a Docker container, or a Virtual Machine (VM) inside the physical machine, up to two of them using a unique public IP address. 
Every relay must define a unique combination of IP address and port number in its configuration.
One advantage of deploying co-resident relays as separate processes is allowing an attacker to simply alter the Tor relay configuration to different port numbers and IP addresses for each relay, minimizing the configuration and computational overhead. 
Once the measurement traffic directed towards one of the co-residing relays is detected by the Traffic Classifier, the regular user's traffic is dropped, thus giving the entire bandwidth and computational resources on the physical machine to the bandwidth scanner, leading to an inflation of the measured bandwidth. 
We note that the bandwidth inflation strategy based on dropping the user traffic was previously proposed and evaluated in \cite{thill2014hidden,johnson2017peerflow}. Previous works, however, focused on individual relays and did not consider scenarios with clusters of co-resident relays. Hence, beyond known attack techniques, our attacker can further inflate the bandwidth by dedicating resources of the entire cluster, not just of the single relay server, to the measured relay. This implies that the inflation factor can be further \emph{amplified by the number of co-resident relays}.  

While it can potentially happen that two co-residing relays are co-measured at the same time, we show in our analysis presented later in \cref{sub:inflation_factor} that this does not happen very often.  With only six BAs responsible for measuring about 7000 relays, and despite the fact that each BA runs 4-5 parallel measurement threads, the probability of such a co-measurement is insignificant, with up to 30 co-residing relays in one cluster. 

\noindent\textbf{Attack Instantiation.} 
An attacker can instantiate the Traffic Classifier (see~\cref{fig:attack_1}) either as an IP filtering system or a machine learning technique. 
However, information about the BAs, including their current IPv4 and IPv6 addresses, is available publicly at Tor Metrics platform~\cite{tor_DAs}. To detect two-hop measurement traffic originating from BAs, we utilize a filter based on BA IP addresses obtained from the same platform. This filter, implemented using Netfilters~\cite{netfilter}, allows us to distinguish such measurement traffic right from the time of arrival of its very first packet and drop user traffic during measurement.

We have various choices available for executing the relay deployment, including the usage of processes, containers, or virtual machines. Given the increased computational requirements associated with containers and virtual machines, we opt to utilize processes for launching our attack.
For each relay, we create separate configurations and start a Tor process based on the corresponding configuration file. To ensure proper functioning, each Tor relay configuration requires a distinct combination of IP address and port number when employing processes.
In contrast, an attacker could leverage Docker containers to eliminate the necessity of specifying these details in the configuration files and instead adopt a more dynamic approach utilizing the Docker ecosystem.

\subsection{\desys~- \sys{} with Dedicated Server}
\label{sub:attack2_design}

\begin{figure}[h]
\vspace*{-0.275cm}
\centering
\includegraphics[width=0.85\linewidth]{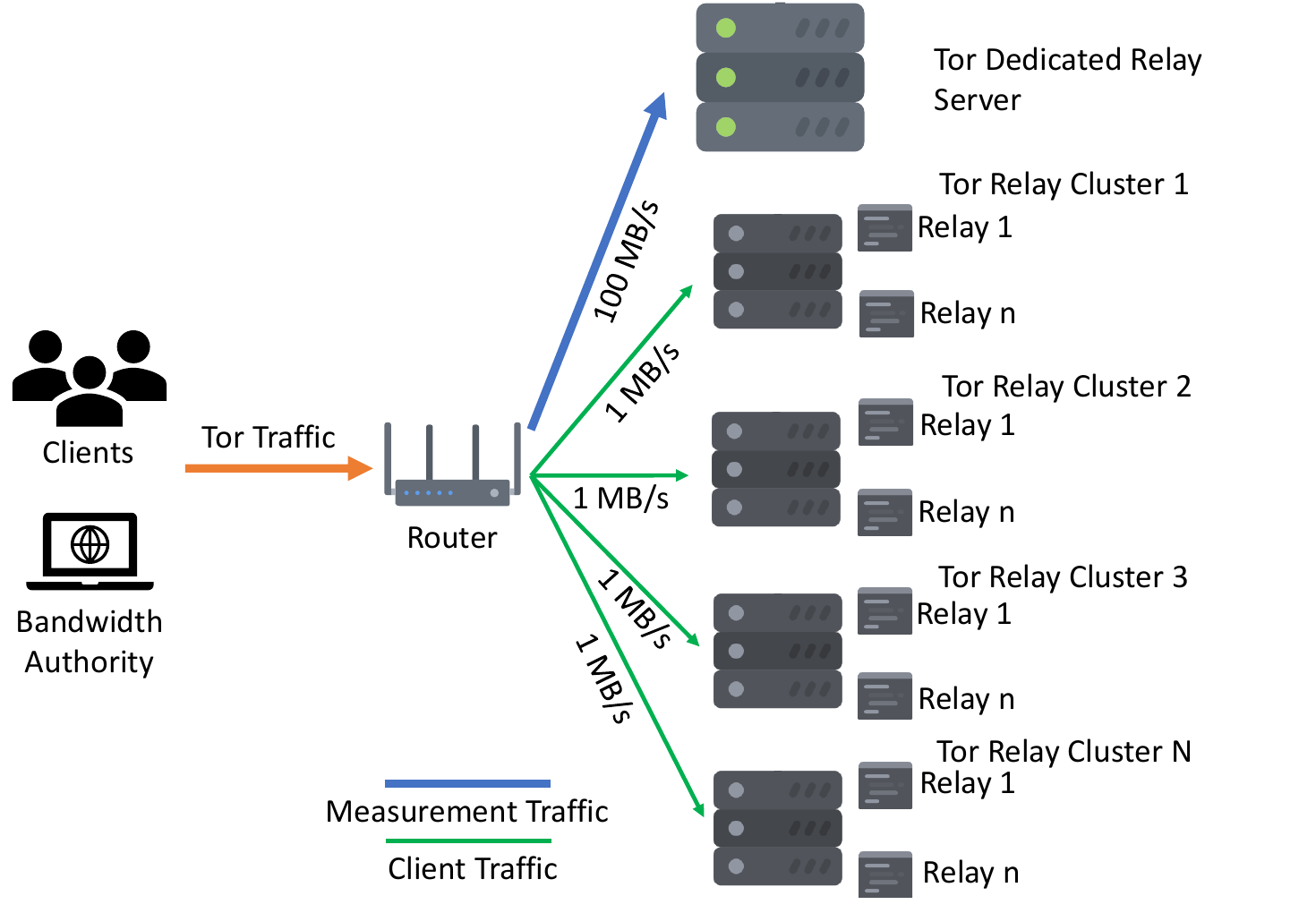}
\caption{\desys: \sys{} with Dedicated Master Server}
\label{fig:attack_2}
\vspace*{-0.375cm}
\end{figure}

\vspace{-0.20cm}
\noindent\textbf{Design.} In the \desys{} scenario, the attacker has control over a network segment and utilizes a physical router or switch to filter and manipulate traffic (see~\cref{fig:attack_2}). This is different from the \cosys{} attack, where traffic filtering and routing functionality is co-resident with the Tor relays (i.e., happens inside the same physical machine). 
As shown in~\cref{fig:attack_2}, the attacker allocates a high-resource dedicated server to handle measurement traffic, while low-resource servers (virtual machines or cost-effective physical machines) are used to host a Tor Relay Cluster and to handle regular user traffic. Each Tor Relay Cluster can accommodate its own cluster of co-resident relays, similar to the \cosys{} attack. 

When the attacker-controlled router or switch detects measurement traffic, it forwards the traffic to the dedicated server. This allows all the attacker-controlled relays to falsely claim the total available bandwidth allocated to the dedicated server, similar to the behavior in \cosys{}. Meanwhile, the relay clusters could continue serving client Tor traffic, e.g., for the purpose of intercepting it and conducting further de-anonymization attacks without much impact on the measurement traffic. 

\noindent\textbf{Attack Instantiation.}
The attack scenario of \desys{} can be instantiated using physical servers and routers. However, the cost of such a setup would be substantial due to the need for a router with multiple public IPv4 addresses. Hence, such a setup is generally more suitable for state-sponsored attackers or other powerful adversaries. 
For cost-efficiency reasons, we instantiated a virtualized version, where the router was deployed using a VM. 
Each relay cluster VM is connected to the virtualized router using OpenVPN connections. As such, we also refer to this attack instantiation as "tunneled" version.
In contrast to the instantiation of~\cosys{}, the relays in~\desys{} are running in Docker containers.
We chose containers since the orchestration of many relays is easier using descriptive definitions via Platform as a Service (PaaS) tools. 
\section{Evaluation} 
\label{sec:evaluation}
In this section, we will begin by outlining the ethical considerations we adhered to during our evaluation. Next, we will present the evaluation environment for our attacks. 
Following that, we will present the evaluation results for both \cosys{} and \desys{}. These results will demonstrate the impact and success of the attacks in terms of inflating the measured bandwidth of the attacker-controlled relays. 
In addition, we will provide a theoretical analysis of the diminishing impact on the inflation factor that can occur when two or more Bandwidth Authorities (BAs) simultaneously measure the bandwidth of the same attacker-controlled relay. Finally, we determine the required amount of attack resources to gain control over a specific part of the Tor Network.

\subsection{Ethical Considerations}
\vspace{-0.20cm}
We initiated the process of responsible disclosure with the Tor team by submitting a bug report via Tor's bug bounty program on HackerOne~\cite{tor-bugbounty-hackerone}.
Additionally, we contacted the Tor Safety Board~\cite{tor-safety-board} to request permission to conduct our experiments on the live Tor network. Unfortunately, our request was declined due to concerns about potential disruptions to user traffic and the quality of service provided by the Tor network. As a result, we were recommended to perform our experiments with special Tor experimentation tools (i.e., a private Tor network), which we precisely did, as described in Sections \ref{sub:attack1_design} and \ref{sub:attack2_design}. Our objective was to ensure the ethical conduct of our research while contributing to the enhancement of the Tor network's security and privacy capabilities.

Furthermore, in the experiments detailed in Section~\ref{traffic_classification}, we gathered both self-generated measurement traffic and client traffic from the real Tor network. It is important to emphasize that when generating measurement traffic, we strictly adhered to the guidelines provided by the Tor Safety Board. Specifically, we scheduled the measurements at different time intervals and utilized various guard and exit relays to prevent any potential denial-of-service (DoS) issues. Regarding the collection of user data, we meticulously designed our process to fully comply with the European General Data Protection Regulation (GDPR) requirements~\cite{GDPR}. We particularly focused on the principle of data minimization (Article 5c) and the regulations for collecting data for scientific purposes (Article 89). As a result, the user traffic we collect is limited to the relay node positioned in the middle, which means the captured packets contain information about the guard and exit relays but do not reveal any details about the users or the specific services they accessed. Additionally, we exclusively capture packet headers such as TCP/IP headers, including TCP high-precision timestamps, while disregarding payloads that carry (encrypted) data.

\subsection{Evaluation Environment}
\label{sub:env_eval}
\begin{figure}[h]
\includegraphics[width=\linewidth]{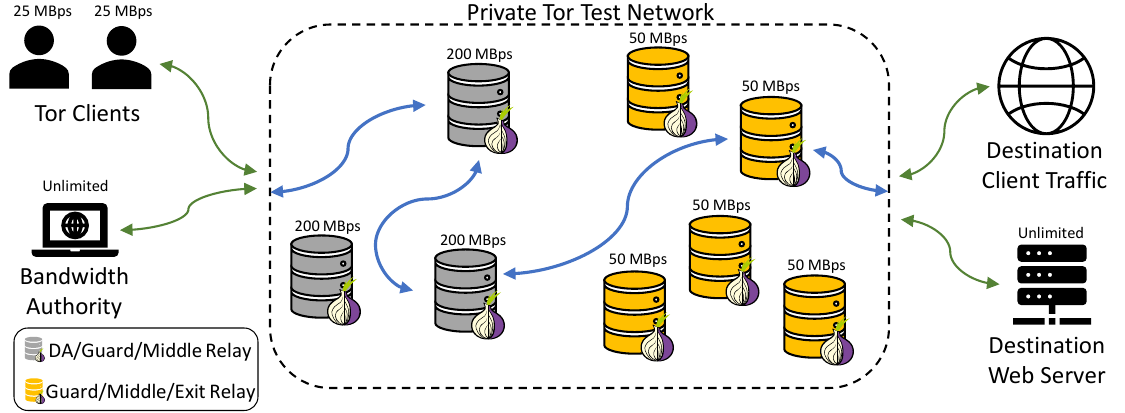}
\centering
\vspace*{-0.375cm}
\caption{Evaluation environment instantiated on the host.}
\label{fig:eval-envt}
\vspace*{-0.375cm}
\end{figure}

\vspace{-0.20cm}
Our experimental setup includes a host machine running Debian Bullseye with 32/64 Cores/Threads, 126 GB of RAM, and 3.2 TB NVMe disk space. 
We instantiate our evaluation environment on this host by running our own private Tor test network.
This environment is used for both attack evaluations.
First, we create the Tor configuration files for a private Tor test network using Chutney~\cite{chutney}, which allows us to create the Tor configuration files based on our input.
Chutney can run a private Tor test network as processes. However, this won't be sufficient for us to instantiate the scenario where (only) co-resident Tor relays share computational resources. Hence, we use Chutney only for bootstrapping the Tor test network configuration. Our Tor test network comprises three DAs, five exit relays, and two clients (see \cref{fig:eval-envt}). Additionally, the evaluation environment contains a BA, a web server, and a connection to the internet. Furthermore, we have the flexibility to dynamically add relays to the private Tor test network even after the evaluation environment has been established. 

After the bootstrap via Chutney, these configuration files are used to create a virtualized environment using a hypervisor.
Here, we use Vagrant~\cite{vagrant} as an orchestration tool and VirtualBox~\cite{virtualbox} as the targeted hypervisor.
Both (Vagrant and VirtualBox) allow us to:
(1) Emulate resources of a server with VMs (not directly possible using Chutney). 
(2) Create a network environment with bandwidth limitations.
A VM is a virtualized physical platform that can run a single relay, a dedicated server, or a cluster of co-residing relays.
The DAs, exit relays, and client relays, previously defined, all operate within their respective VMs.
Each VM has two vCPUs and 4 GB RAM if not otherwise declared.

Since we conduct our \sys{} attacks in a virtual setting with a Tor test network, the inflation factor must be computed the same way as in the real Tor network. 
As a result, we deploy an SBWS instance for bandwidth measurement and report the mean bandwidth published by SBWS.
Thus, we have a separate VM for the BA (using SBWS) and another VM serving as the destination web server for measuring purposes.
To control the network bandwidth for the VMs, we make use of VirtualBox's bandwidth limitation mechanism. Specifically, we set the bandwidth limit to 200 \emph{MBps} for the DAs, 50 \emph{MBps} for the relay VMs, 25 \emph{MBps} for the clients, and leave the BA and the web server with unlimited bandwidth.

Once the VMs are instantiated, we utilize Ansible~\cite{ansible_git} playbooks to orchestrate the installation of the Tor package \cite{relayops} and include the configuration files. Furthermore, the SBWS service is configured to use our target web server as a measurement destination. This enables us to measure each relay's bandwidth in our evaluation environment. 
Next, we run a VM for each relay (i.e., DAs, BA, clients, exit relays) with a unique IP address and a unique set of ports. 
The most important configuration is the port number, as this value will be advertised for incoming Tor connections. 
The corresponding ports are then opened in the firewall to allow incoming packets. 
Finally, we start each relay as a separate CPU process. 

After setting up the private Tor test network's benign entities, we include malicious entities such as relays performing the~\sys{} attacks and a single malicious relay dropping user traffic. 
We also set identical ``MyFamily'' values for malicious relays of our~\sys{} attacks by adding the identity fingerprints of each relay so that the clients avoid using more than one of them in the same circuit (i.e., avoid using middle and guard relays in the same circuit), thereby avoiding an overload on our server due to `short circuits'.

To introduce traffic in our network, the clients request the download of random files. To establish a baseline, we measure the provided bandwidth of our test networking using the SBWS service without an ongoing \sys{} attack (e.g., only a single malicious relay dropping user traffic is deployed, but no clusters). 
The baseline measurement indicated a bandwidth of approximately 25 \emph{MBps}. This value served as the starting point for our subsequent evaluations and comparisons.

\subsection{Evaluation of \cosys{}}
\label{sub:attack1_eval}
\vspace{-0.20cm}
In our implementation of \cosys{}, we deployed five malicious relays in a single VM to assess their effectiveness within our evaluation environment. Each of these relays was executed as a separate process for ease of deployment. It's important to note that the choice of five relays was determined by the computational limits assigned to the VMs used in our setup, and this number can be adjusted accordingly. 
Additionally, we conducted tests using a deployment strategy in which co-resident relays were introduced to the network at staggered intervals, ranging from a few hours to a few days. This approach was implemented to minimize the likelihood of co-resident relays being simultaneously measured in the actual Tor network. 
The attack is launched once all deployed relays are connected to our Tor test network.
We activated our IP-based traffic detection mechanism (as outlined in \cref{sub:advmodel}) to distinguish measurement traffic from the BA and drop user traffic during measurement. 

\begin{figure}[htbp]
    \vspace*{-0.3cm}
    \centering
    \includegraphics[width=0.7\linewidth]{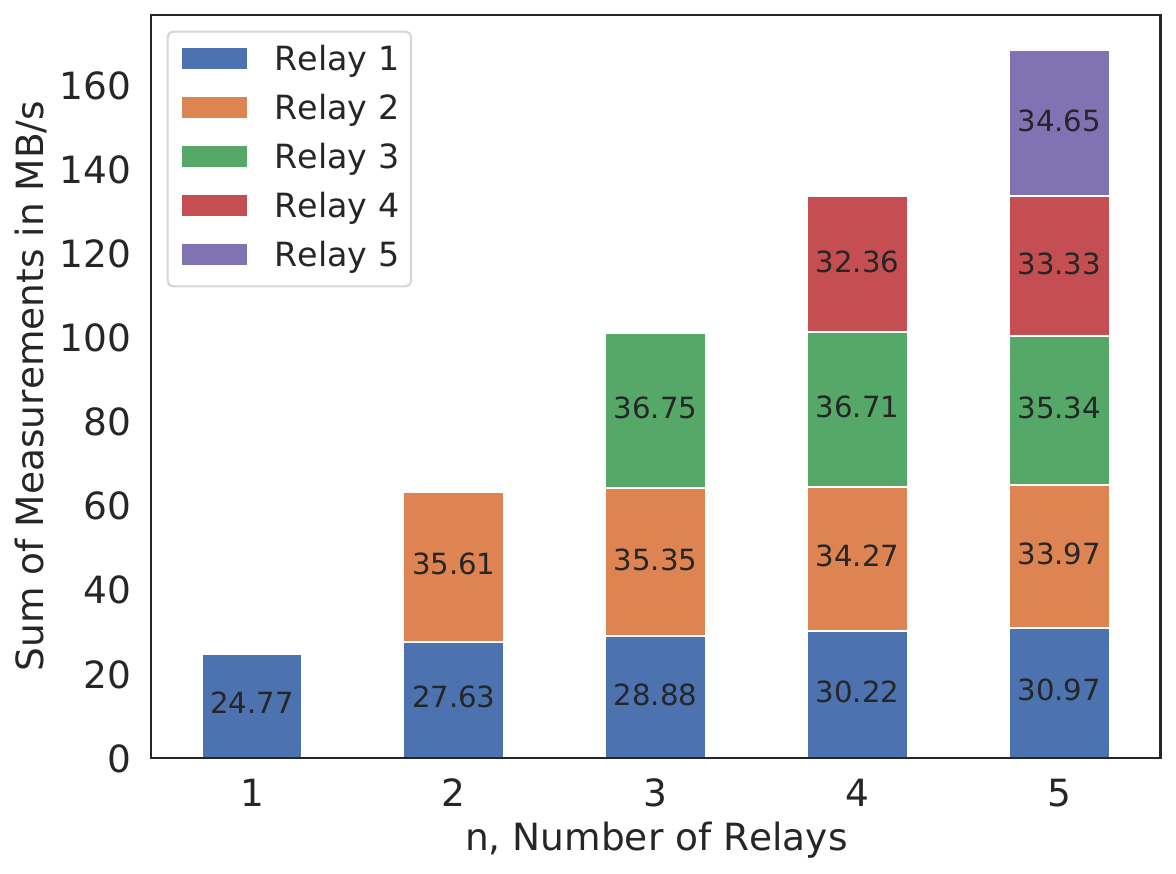}
    \vspace{-0.3cm}
    \caption{\cosys{} - Measurements for 5 co-resident relays}
    \label{fig:measurement_results_5co}
    \vspace*{-0.45cm}
\end{figure}

We observed that with the addition of any new relay, the average of the measurement reported by SBWS after at least three measurements slightly changes, as depicted in \cref{fig:measurement_results_5co}.
For example, the average measurements of the `blue' relay (Relay 1) steadily increase.
We have noticed that the measurements obtained from new relays surpass the established baseline. Upon investigation, we discovered that the baseline measurements incorporate some 'unlucky' measurements. A similar observation can be made regarding the measurements of \cosys{}. 
Slight changes are caused by the encryption/decryption and the overhead introduced by packet dropping that occurs until user connections are closed. 
Further, exit relays involved in the measurement circuit may also be used by other circuits in our Tor test network, which can potentially reduce the impact of the attack, e.g., measured bandwidth. 
As seen from \cref{fig:measurement_results_5co}, \cosys{} attack strategy allows an attacker to boost an inflation attack by a factor above $n$ linearly due to variance in the measurements. 

\subsection{Evaluation of \desys{}}
\label{sub:attack2_eval}
\vspace{-0.20cm}
For \desys{}, we instantiate three relay clusters as VMs (with a bandwidth limit of 25 \emph{MBps}), each hosting and carrying a cluster of six Tor relays as Docker containers in the same physical platform as the virtual router (also a VM with 200 \emph{MBps} bandwidth limit).
Further, we add a dedicated relay server with 6 vCPUs, 12 GB RAM, and apply a bandwidth limit of 50 \emph{MBps}.
The rest of the evaluation environment is the same as in our experiments for~\cosys{}. 

\begin{figure}[h]
    \centering
    \includegraphics[width=0.7\linewidth]{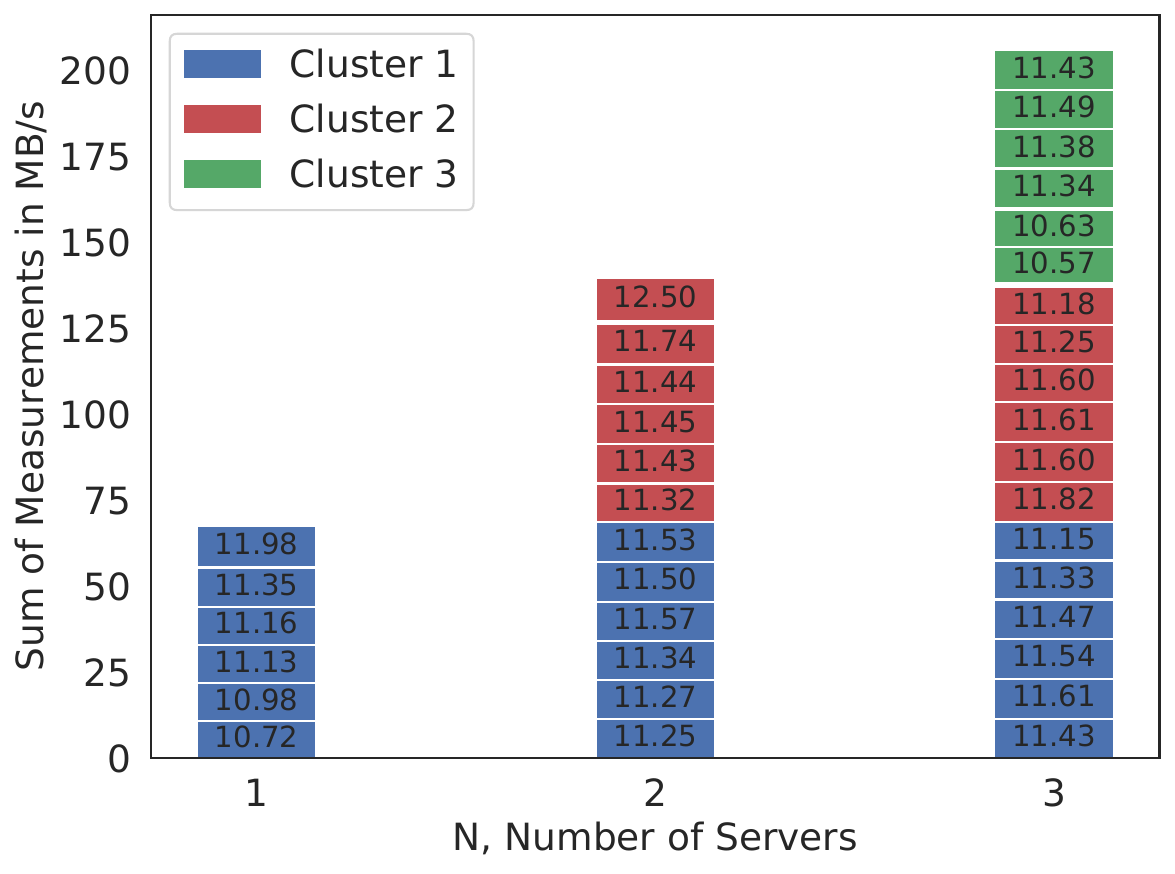}
    \vspace*{-0.275cm}
    \caption{\desys{} - Measurements for 3 clusters with 6 relays each.}
    \label{fig:attack_2_results}
    \vspace*{-0.375cm}
\end{figure}

We generate continuous user traffic to each of the six relays deployed in the three relay cluster VMs.
We simulated the behavior of the BAs by running an instance of SBWS to measure the bandwidth of the relays.
As a result of SBWS measurements, each relay claimed more than a fifth (10 \emph{MBps}) of the virtual server's bandwidth, which is reduced from 50 \emph{MBps} due to the VPN, routing, and Tor's overhead. 
As shown in \cref{fig:attack_2_results}, the attacker claimed over 65 \emph{MBps} for all six relays in Cluster 1. Just by adding two additional clusters to the VPN router, an attacker can achieve an impressive 204.43 \emph{MBps} with 16 relays in three clusters.

The results demonstrate that the \desys{} attack variation exhibits a linear relationship with the number of relays $n$ in each cluster, which are distributed across $N$ relay cluster servers. As a result, the attacker has the ability to amplify the inflation by a factor of nearly half $n*N$ by combining our attack with any other inflation attack. This means that the attack's impact can be significantly increased by scaling up the number of attacker-controlled relays inside a cluster and increasing the number of attack servers (i.e., clusters). 

\noindent\textbf{Comparison of~\cosys{} and \desys{}.} When comparing \cosys{} and \desys{} attack versions, it is evident that the routing and VPN overhead significantly affect the available bandwidth in~\desys{}. Nonetheless, the practical expenses associated with deploying numerous servers on the real Tor network are higher when employing~\cosys{}. Conversely,~\desys{} offers a more cost-effective alternative, enabling the attacker to acquire a small server and multiple IPs. 

\subsection{Exploring the Inflation Factor}
\label{sub:inflation_factor}
\vspace{-0.20cm}
The presence of multiple relays in an attacker's co-resident cluster increases the likelihood of simultaneous measurements, also known as "coincidence," occurring on multiple relays. This coincidence phenomenon reduces the ability of these co-resident relays to falsely claim the total available bandwidth on the co-resident server, resulting in a decrease in the inflation factor. This effect didn't manifest in experiments we conducted in \cref{sub:attack1_eval} and \cref{sub:attack2_eval} due to the small sizes of attacker-controlled clusters.

The coincidence rate refers to the probability of measurements occurring simultaneously on more than one co-residing relay within the cluster. It quantifies the frequency at which multiple relays in the cluster experience measurement events at the same time, which impacts the effectiveness of the inflation attack.
By calculating the coincidence rate, an attacker can make informed decisions regarding the number of co-resident relays to deploy in order to maximize their expected bandwidth inflation gain or inflation factor.

We conducted an analysis of the coincidence rate in the real Tor network by examining the published bandwidth files that contain measurement information from the Bandwidth Authorities (BAs). The objective of this analysis was to determine the number of relays from a selected group that were co-measured, meaning they were measured simultaneously.

A challenge in this analysis is that the bandwidth files provided by the BAs only include a single timestamp, which indicates the end of a measurement. However, to calculate co-measurements, it is crucial to have information about the duration of the measurement, which can only be determined by knowing the start timestamp as well. To address this gap in knowledge, we employed a method of creating speculative realities or simulations. In these simulations, we randomly assigned start timestamps while adhering to known constraints such as the minimum duration of the measurement. We then analyzed the results from multiple simulations and calculated the average values of measurement durations. 
This approach allowed us to estimate the coincidence rate by simulating different measurement scenarios and obtaining insights into the likelihood of relays being measured simultaneously. By conducting this analysis, we gained valuable information about the coincidence rate in the Tor network and its impact on the effectiveness of co-resident relay attacks.

\noindent\textbf{Measurement Duration.}
To analyze the duration of measurements in the collected bandwidth files, we conducted the following steps: First, we obtained the bandwidth files from all Bandwidth Authorities (BAs) for the months of May, June, and July in 2022 from CollecTor \cite{collector}. This dataset consisted of 10,440 files from 2,208 measurement rounds, providing a comprehensive view of the measurements conducted during that period. Due to the downtime of the "moria1" BA since April 2022 and intermittent downtime of other BAs, we observed a limited number of bandwidth files per hour during data collection. On average, we had only four or five bandwidth files available per hour. Second, we extracted the fingerprints of the measured relays for each hour using the Stem library. The fingerprint uniquely identifies a relay in the Tor network and allows us to track its measurements over time. Third, since the bandwidth files only provide the ending timestamp of a measurement, we needed to infer the duration of the measurements. We relied on the information provided in the SBWS documentation, which states that a single measurement takes at least 5 seconds but no more than 10 seconds. Based on observations, where measurements from the same BA were found to be less than 25 seconds apart, we have accounted for BAs employing multiple processes or threads for measurement. To gain further insights into measurement duration, we randomly assigned each measurement a thread identifier, ranging from 0 to $n$, based on the ordering of the measurements by their ending timestamp. We calculated the duration of each measurement by subtracting the ending timestamp of the $(i-1)$-th measurement from the $i$-th measurement. 
In cases where an existing identifier did not satisfy the 25-second constraint (the minimum measurement duration), we incrementally created a new identifier $(n+1)$. 
Since there are numerous possible combinations of thread identifier arrangements, we repeated the random assignment process for 120 iterations.

In the majority of iterations (88.37\%), the measurements were assigned four unique threads, fulfilling the 25-second constraint. In 99.49\% of the random assignment iterations, the BAs utilized between one and six threads for their measurements. We considered measurements that were less than 50 seconds apart in these timelines to be executed sequentially without failed measurements in between. This assumption is based on the observation that the duration of a measurement can be calculated by subtracting the end timestamp of the previous measurement ($i-1$) from the end timestamp of the current measurement ($i$). If there are failed measurements in between, the resulting duration would exceed 50 seconds, indicating that the two measurements were not executed sequentially.
For those measurements meeting the sequential execution criteria, we calculated a median measurement duration of 39 seconds. This calculation provides an estimation of the typical duration for measurements conducted by the BAs, considering the constraints and observations described above.

\noindent\textbf{Coincidence Rate.}
For our experiment, we focused on a specific group of relays within the Tor network. After analyzing various families of relays, we decided to work with the "Artikel10" family \cite{artikel}, which consists of 120 relays with a combined total bandwidth of 2069.17 MBps. This family showed indications of co-residing relays, such as multiple pairs of relays sharing the same IP address. Additionally, within the family, there were several sets of relays that exhibited the same up-time and advertised similar bandwidth values. These observations suggested that these relays might be deployed on the same physical machine, exhibiting behavior similar to our \cosys{} attack.

To determine the coincidence rate, we utilized the insights gained earlier regarding measurement duration. We assigned each measurement a start timestamp that occurred 39 seconds before the measurement's end timestamp. By doing this, we simulated the duration of the measurements.
To investigate the coincidence rate, we selected a random subset of $N$ relays from the ``Artikel10'' family and filtered the bandwidth files to include only the relays present in our chosen set. We then examined the measurements within the filtered dataset and identified those with overlapping ranges. An overlapping range indicates that another measurement started prior to the current one and was still ongoing, or that a measurement started during the current measurement.

We proceeded to count the occurrence of specific events, namely, the number of simultaneously measured relays out of the $N$ relays in our set. We varied the value of $n$ from $1$ to $N$ to analyze the incidence of simultaneous measurements across different numbers of relays. 

\begin{figure}[htp]
    \centering
    \includegraphics[width=\linewidth]{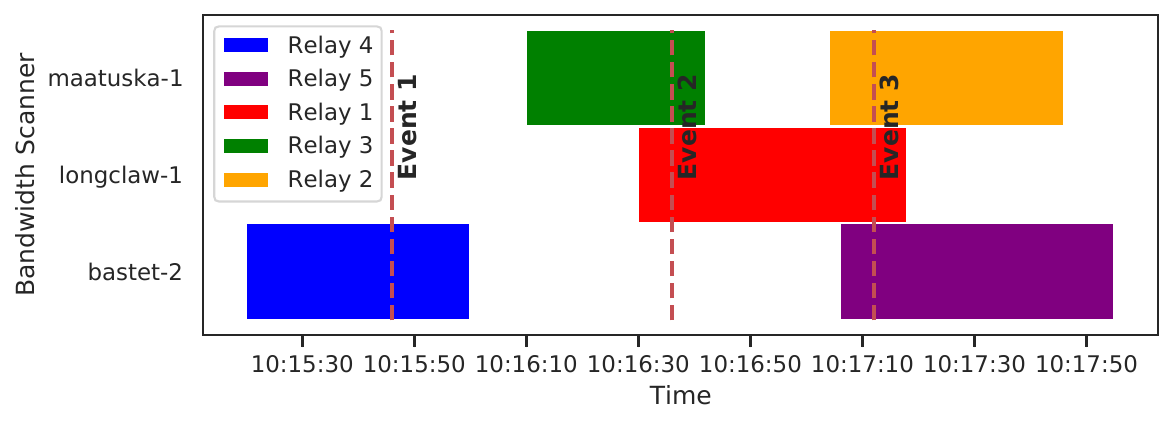}
    \caption{An example of measurement events.}
    \label{fig:overlap}
    \vspace*{-0.375cm}
\end{figure}

In Figure \ref{fig:overlap}, we illustrate an example of how the counting of measurement events is performed. The timeline represents the measurement periods for a group of five relays, where three different events occur. In the first event (Event 1), Relay 4 (blue) is measured separately without any overlap with other relays. The second event (Event 2) shows an overlap between the measurements of Relay 1 (red) and Relay 3 (green). This overlap results in two out of the five relays being counted as co-measured since their measurement periods coincide. Similarly, the third event (Event 3) exhibits an overlap among three out of the five relays, namely Relay 1, Relay 3, and Relay 5.

Although short periods of overlap, even less than one second, may cause a minor delay in measurement results, we counted all the simultaneous measurements regardless of the overlap duration. To calculate the probability of each event, we compute the ratio of its occurrence to the total number of measurements within the three-month time period. This allows us to determine the likelihood of each event happening relative to the overall measurements conducted during that period.

\begin{figure*}[htbp]
\begin{subfigure}[b]{0.30\linewidth}
    \includegraphics[width=\linewidth]{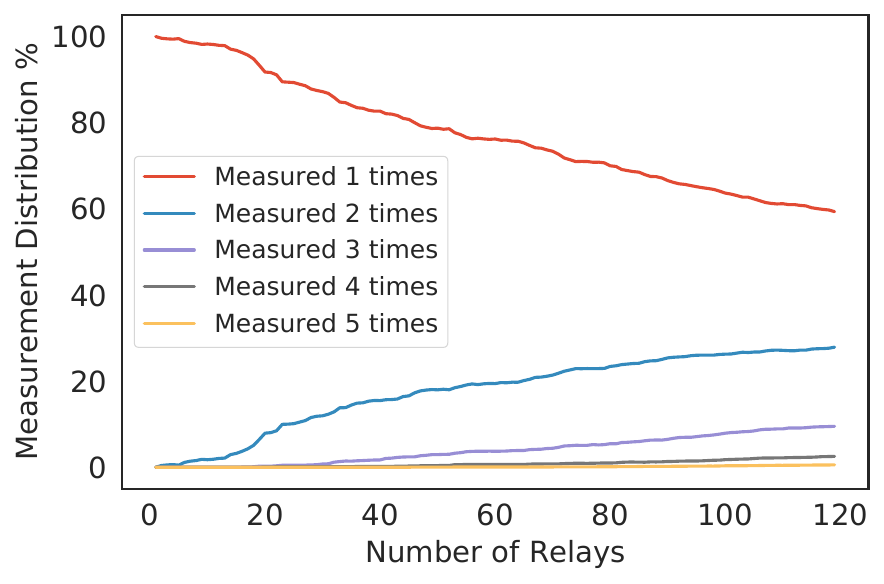}
    \vspace*{-0.675cm}
    \caption{}
    \label{fig:simulationMeasurementDistribution}
\end{subfigure}%
\hfill%
\begin{subfigure}[b]{0.30\linewidth}
    \includegraphics[width=\linewidth]{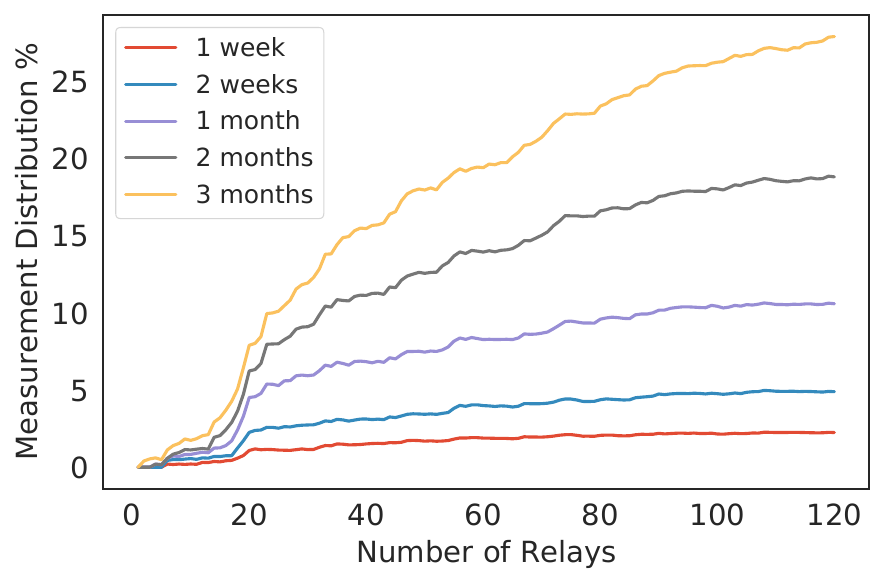}
    \vspace*{-0.675cm}
    \caption{}
    \label{fig:simulationMeasurementTimePeriods}
\end{subfigure}
\hfill%
\begin{subfigure}[b]{0.30\linewidth}
    \includegraphics[width=\linewidth]{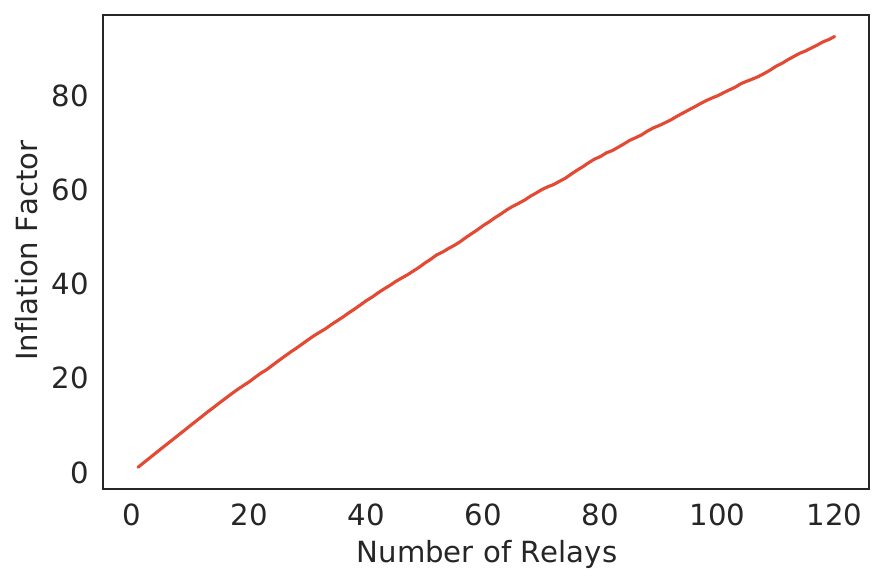}
    \vspace*{-0.675cm}
    \caption{}
    \label{fig:simulationInflationFactor}
\end{subfigure}
\vspace*{-0.375cm}
\caption{Measurement distribution (a) for a simulation with a 3-month time window, (b) for two relays being measured at the same time for different time windows and cluster sizes. (c) Theoretical inflation factor per relay for no. of relays.}
\vspace*{-0.375cm}
\end{figure*}

\noindent\textbf{Measurement Distribution.} In~\cref{fig:simulationMeasurementDistribution}, we present the distribution of measurements among the selected relay set. We observed that the occurrence of three or more relays being co-measured is rare when the number of relays ranges from 1 to 30. 
However, as the number of relays increases, the probability of two relays being co-measured also increases. The probability reaches its peak at 27.91\% when there are 120 relays in the set. Nonetheless, the majority of relays are still measured individually, accounting for 59.17\% of the measurements. This allows them to claim the full bandwidth of the dedicated server.

To recall, we defined the cluster size as the number of relays per dedicated server. 
For a cluster size of 45 relays, the probability of three relays from the set being co-measured is only 0.014\%. Even with 120 relays, the probability remains very low at 0.15\%. These low probabilities indicate that the occurrence of three or more relays being co-measured is a rare event.

In our study, we also investigate the impact of the number of relays in the set and the duration of the simulation on the distribution of co-measurements. We focus our temporal analysis on the event of two relays being co-measured, as this event is more likely to occur compared to three or four relays being co-measured, as shown in~\cref{fig:simulationMeasurementDistribution}. By varying the number of relays in the set and the duration of the simulation, we can observe how the probability of two relays being co-measured changes over time. This analysis provides insights into the likelihood of such coincidences occurring and helps us understand the influence of different factors on co-measurements.

The results depicted in~\cref{fig:simulationMeasurementTimePeriods} illustrate the outcomes for different time periods and sets of varying sizes. One noteworthy observation is that longer simulation periods tend to yield more occurrences of simultaneous measurements. For instance, in the case of a set containing 120 relays, the probability of two relays being co-measured is 12 times higher over a span of three months compared to just one week. This finding suggests that, from an attacker's perspective, considering the simulation results within a three-month timeframe would be more appropriate as it represents the worst-case scenario. It is important to note that results may differ in the context of an even more extended attack period.

\noindent\textbf{Inflation Factor.} 
The theoretical analysis of the inflation factor for our attack, considering the worst-case scenario of a three-month period, provides insights into the attack's potential based on real measurement data. However, it is important to note that the actual inflation factor may be affected by factors such as packet overhead and network latency.

In our analysis, we take into account that relays measured by a single BA can claim the full bandwidth of the dedicated server, while relays co-measured by two or more BAs must evenly share the available bandwidth. We calculate the expected inflation factor for different numbers of relays, assuming a contribution factor of one for a single relay and a contribution factor of 0.5 for relays co-measured by two BAs.

As illustrated in~\cref{fig:simulationInflationFactor}, the inflation factor gradually decreases as the number of relays increases. For small cluster sizes, the attack demonstrates nearly linear inflation rates. For example, with a cluster of 10 relays, the inflation factor reaches 9.91. This value nearly doubles to 19.16 when using 20 relays, and employing 30 relays results in a nearly tripled inflation factor of 28.03. These findings highlight the potential for significant bandwidth inflation with a relatively small cluster of co-resident relays.

However, as the number of relays in the set increases, the inflation factor no longer exhibits a linear relationship. For example, with 120 relays, the inflation factor is 92.52, indicating that approximately 28 relays are not actively contributing to the attack due to the higher rate of co-measurements in larger cluster sizes.

To summarize, in sets with a small number of relays, individual relays are more likely to be measured independently, allowing them to claim the full bandwidth of the dedicated server. Only a few cases show three or more relays from the set being co-measured simultaneously. Theoretical analysis demonstrates that the attack's capacity shows almost linear growth when employing a small number of relays. However, as the number of relays in the set increases, the attack's effectiveness diminishes, leading to a decrease in the inflation factor.

\subsection{Estimation of Required Attack Resources}
\label{sub:opt_resources}
\vspace{-0.20cm}
We used a curve fitting algorithm to create a curve that matches the inflation factor data (cf.~\cref{eq:inflation_factor} in Appendix). 
This curve allows for estimating the inflation factor for a given cluster size.

As an example, we calculated how many resources an adversary needs to control half of the Tor network traffic. On average, the Tor Network had an advertised bandwidth of around 678 GBit/s in the year 2022~\cite{collector}. Thus, we derive that, to control half of the traffic an adversary needs to advertise an additional 678 GBit/s to the network which can be achieved with 1090 relays (cluster size of 109) and ten dedicated servers, each providing 100MB/s. In summary, gaining control over half of the Tor network's traffic only requires a little effort and can be accomplished by utilizing 10 dedicated servers with a bandwidth of 100MB/s running the \sys{} attack. 
\section{Additional Insights on Tor}
\label{sec:insights}
In this section, we provide some noteworthy insights we gained while performing certain analyses on the Tor network.

\noindent\textbf{Existence of Relay Families.}
During our preliminary investigation of relays in the Tor network, we came across another interesting observation $-$
\emph{co-resident} relays, i.e., multiple relays deployed and operating within the same physical or virtual machine. The main identifiable attributes of such co-resident relays that we observed included, sharing of the same public IP address, same uptime and similar measurement results by multiple relays.
Taking these identifiers into consideration, we closely examined relays located in Germany and were able to identify several families with potentially co-residing relays (see \cref{tab:families}). 
The largest family we observed consisted of 178 relays controlled by the same operator, while the most powerful (in terms of offered bandwidth) family observed, offered nearly 2600 \emph{Mbps}. 
It is also important to highlight that these identifiable attributes were also found in families with number of relays as low as 3. 
Another important thing to note here is that it is possible for relays in such families (of co-resident relays) to claim the entire available bandwidth if each of them is measured separately. 
Evidence of the presence of families of co-resident relays controlled by the same entity in the Tor network, coupled with the fact that it is possible for relays in such families to claim the entire available bandwidth when measured by a BA, is a bit concerning as it could expose the Tor network to a potentially new form of inflation attack, as proposed by us in this work.
A recent discussion in the Tor community suggests a proposal to change the two relays per IP limitation for relay operators either by a special request (by relay operator) or a general increase of 32 relays per IP address \cite{increase_relays_ip}. The main reasoning behind this increase is to make Tor more scalable in modern CPUs with a high number of cores and if successful, this will further enhance our \sys{} attacks.

\noindent\textbf{Port Usage by Tor Relays.}
Before performing our analysis on the real Tor network, we also investigated the ports that could be used for our relays and observed that the relays use more than 581 different ports (see \cref{fig:port_usage}). 
Further, this could also potentially be a sign of co-residing relays operated by the same entities on Tor.
Tor operators are free to set the port as per their convenience based on their firewall and routing policy \cite{port_setting}. 
Therefore, deploying \sys{} co-residing relays with a different set of ports will not be considered suspicious.

\section{Resilience of Alternate Bandwidth Measurement Solutions}
\label{sec:res_alt_solutions}
In this section, we discuss the resilience of alternative bandwidth measurement solutions against \sys{} attacks. 

\subsection{3-Hop Measurement Circuits}
\label{traffic_classification}
\vspace{-0.20cm}
Greubel et al. \cite{andre2018smartor} already pointed out that 2-hop measurement circuits in Tor can be easily detected using BA's IP addresses. As a countermeasure, the authors proposed to utilize 3-hop circuits during measurements and place the measured relay in the middle position (see \cref{fig:3hop}), in between randomly chosen guard and exit relays, thus obscuring the source and destination of the traffic. One needs to note, however, that such a solution puts an additional burden on the Tor network since more relays will be involved in carrying measurement traffic. 

However, given the rise of effective AI methods, it is questionable if such an approach can withstand modern adversaries and if the increased load on the network is justified. To this end, we evaluate the resilience of the 3-hop method by (i)~collecting the dataset for training in a 3-hop circuit setting, and by (ii)~building an ML model based on a traffic classification method that utilizes a Self-Attention Convolutional Neural Network (SA-CNN) proposed by Guorui et al. \cite{SAMNNModel}.

\noindent\textbf{Data Collection.} In order to train a supervised model such as SA-CNN, we collect Tor network data and label it as: (i) client traffic data, and (ii) measurement traffic data.  

\noindent\emph{Collecting Measurement Traffic.} To collect measurement traffic, we first deployed our own \emph{measurement} relay node within the Tor network and ensured that it reached phase four of its life cycle, i.e., the steady-state guard relay (reached after 68+ days) \cite{TorLifeCycle2013}. 
Then, by means of the Python-based Tor controller library \emph{Stem} \cite{stem_library}, we instantiated our own BA service which was modified to build 3-hop circuits for measurements. The measurement relay was placed in the middle position, while the guard and exit relays were randomly chosen among candidates that (supposedly) offered higher bandwidth than the measurement relay (to eliminate bottlenecks). Additionally, the measurement relay was also used for traffic collection. 
We then used the 3-hop circuit to download four files with the sizes 16, 20, 50, and 100 \emph{MB} from three different servers, mimicking a measurement service of SBWS (as described in \cref{sec:background}). 
Meanwhile, we captured traffic arriving at and leaving the measurement relay using \emph{tcpdump} and filtered out any other Tor traffic. Doing this was straightforward as the time of measurements, as well as the IP addresses of the guard and exit relays built into our measurement circuits were known and under our control. By using this strategy, we captured roughly 3.4 \emph{GB} of measurement traffic.

\noindent\emph{Collecting Client Traffic.} Client traffic was similarly collected on the measurement node of the 3-hop circuit. Client-generated traffic was separated from the self-generated measurement data by filtering out packets arriving from guard relays and addressed to the exit relays. 
We also had to filter out measurement traffic produced by the legitimate Tor measurement service, which was done using the (known) IP addresses of the BAs. Our measured relay was able to capture 118 \emph{GB} of client data in this setting. 

\noindent\textbf{Traffic classification using Machine Learning.} Next, we describe details of the ML model we employ for traffic classification, specifically the Self-Attention Convolutional Neural Network (SA-CNN) proposed by Guorui et al. \cite{SAMNNModel}, including (a)~traffic pre-processing, (b)~feature engineering, (c)~model training, and (d)~model evaluation. 

\noindent\emph{Traffic Pre-processing.} 
To clean up the captured measurement and client data, we removed all packets used for establishing TCP connections, for example, SYN and SYN-ACK, as they are not relevant to measurement traffic classification. We then matched each packet in the captured data to their corresponding network flow based on the source and the destination IP addresses (i.e., grouped packets belonging to the same source and destination combination), with each flow not exceeding 1,000 packets (as per specifications in the SA-CNN model \cite{SAMNNModel}). 
In total, we obtained 747 self-generated measurement flows with 390,276 packets. Using only 20\% of client data, we obtained 1,591,820 client traffic (non-measurement) flows with 12,776,330 packets in total. 
Note that we only used a fraction of the obtained non-measurement data during training to prevent class imbalance in the dataset. 

Following guidelines proposed in SmarTor \cite{andre2018smartor}, we also masked the source and destination IP addresses of the IP header in the packets to simulate a real 3-hop scenario where the guard and exit relays in the circuit are chosen at random, and thus the aforementioned information is not known. 
In addition, we also masked the checksum, identification, offset fields, source, and destination ports of the TCP header, the sequence number, and the ACK flag since, according to Guorui et al. \cite{SAMNNModel}, these confuse the Self Attention CNN model and could result in the model not generalizing well. 

\vspace{0.2cm}
\noindent\emph{Feature Engineering.} 
As our captured TCP data also contains high-precision timestamps, to prevent the machine learning model from classifying the data based on timestamps, we calculated the inter-arrival time of subsequent packets and used that value for analysis. This enables our model to classify traffic data based on statistical features of the network flow. 
Further, as Akbari et al. \cite{akbari2022traffic} pointed out that traffic classification models trained using encrypted TLS payload data also fail to generalize, we omit encrypted payloads from our captured packet dataset to prevent our classification model from learning TLS cipher information. 

\noindent\emph{Final Dataset.} 
We trained our traffic classification model using the labeled dataset comprising of Tor measurement and client traffic data flows, as outlined above.  
The entire dataset was split into train, validation, and test sets with 70\%, 20\%, and 10\% of total measurement flows in each set, respectively. 
The measurement to client traffic packet ratio in the training, validation, and test sets was 52.1\%, 57.8\%, and 60.8\%, respectively. 
We included only a fraction of the captured client traffic data in the training dataset, to keep the classes balanced.

\noindent\emph{Model Building and Training.} We select the Self-Attention Convolutional Neural Network (SA-CNN) proposed by Guorui et al. \cite{SAMNNModel} for traffic classification. Their work is especially suitable for traffic classification needed for inflation attacks since it allows packet-level input and can produce output classification results as packets arrive without needing to observe large portions of traffic flows. Such online (or real-time) classification is a crucial requirement for inflation attacks such as \sys{} due to the need for on-the-fly traffic re-routing to forward 
measurement and client traffic to a pre-determined destination. 
While no major changes and parameter optimizations were required to adapt the SA-CNN model for our problem, the input dimension of the model was changed to a single packet with a total length of 44 \emph{Bytes} (comprising of 20 \emph{Bytes} of IP header, 20 \emph{Bytes} of TCP header, and 4 \emph{Bytes} to store the inter-arrival time calculation).

\begin{figure}[h]
\includegraphics[width=0.8\linewidth]{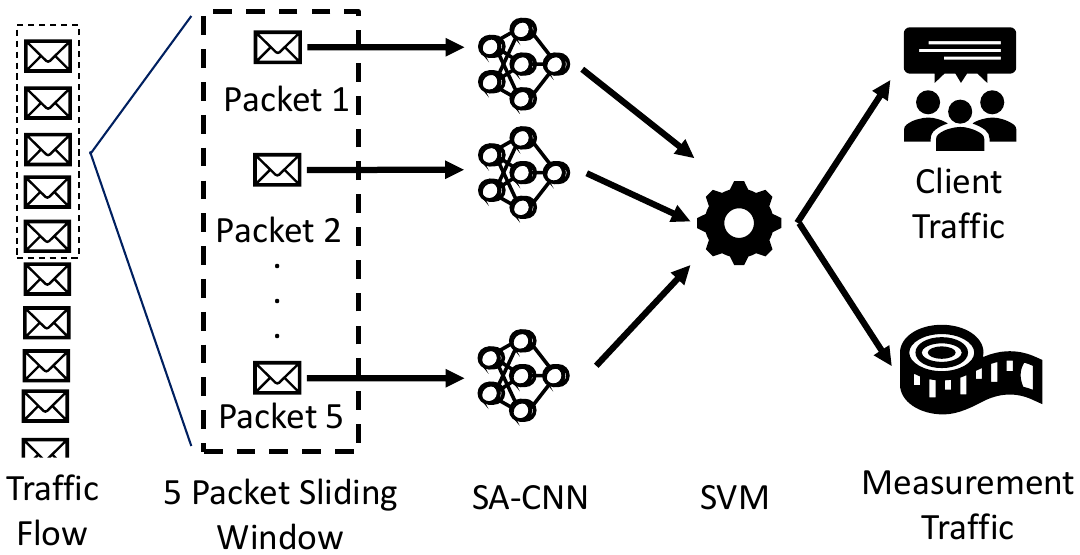}
\centering
\vspace*{-0.2cm}
\caption{Cascaded Model.}
\label{fig:SVMDesign}
\vspace*{-0.375cm}
\end{figure}

Although we were able to achieve high accuracy for both the measurement and client traffic classes when using only 2.6\% of the obtained client traffic in the training dataset, we noticed increased false positives (i.e., client traffic incorrectly classified as measurement traffic) when the model was evaluated on the whole dataset. 
Thus, to mitigate such false positives and improve accuracy, we employ a \emph{cascaded model} where multiple classification outputs from the SA-CNN model are combined and classified using a \emph{Support Vector Machine (SVM)}, as shown in Figure \ref{fig:SVMDesign}. 
Specifically, we use a sliding window (with 80\% overlap) to select windows of size five (packets) from a traffic flow, and each individual packet (in the window) is then input into the SA-CNN model. The binary classification outputs generated by the SA-CNN model for the five packets are then combined and passed on to the SVM as input. For every five packet-based window, the SVM outputs a final binary classification deciding between either a measurement or client traffic class. 
To generate a labeled dataset for the SVM model, we first obtained outputs for only client traffic input data, followed by only measurement traffic from the SA-CNN model. We used 20\% of this new dataset to train the SVM model and 80\% for testing.

\noindent\textbf{Evaluation.} 
We analyze the performance of our above-trained classifier for detecting measurement traffic in 3-hop circuits using the standard performance metrics of precision, recall, and F1-Score. To recall, \emph{precision} (or ratio of correct predictions) indicates how often our classifier is correct when it predicts measurement or client traffic packets.
\emph{Recall} (or sensitivity) is the proportion of measurement and client traffic packets that were correctly identified. 
The \emph{F1-score} is calculated as the harmonic mean between the precision and recall (used as a reliable indicator for model performance, as it also accounts for class imbalance \cite{NNmetricsPowers}).

\begin{figure}[h]
\includegraphics[width=0.8\linewidth]{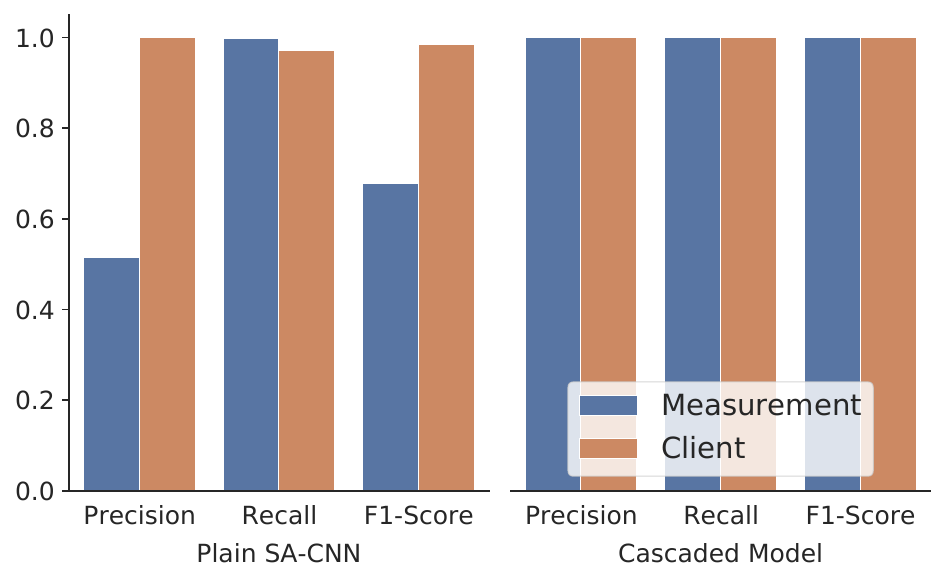}
\centering
\vspace*{-0.2cm}
\caption{Performance comparison for Plain SA-CNN and Cascaded Models.}
\label{fig:modelPerformance2}
\vspace*{-0.375cm}
\end{figure}

We evaluated our proposed \emph{cascaded model} approach, and as seen in \cref{fig:modelPerformance2}, we were able to 
successfully mitigate the impact of false positives and achieve F1-Scores of above 99\% for both classes (compared to the 67\% F-Score for measurement class in Plain SA-CNN model).
Although our cascaded model approach for traffic classification can only be applied once at least five packets in a traffic flow are captured, this introduces only a slight delay and is acceptable as measurement flows typically consist of thousands of packets.

We measured the inference time of our SA-CNN model on an Nvidia A16 (4x 16GB memory) graphics card \cite{nvidia_a16} for different batch sizes (5, 128, 512, 2048, 4098, 8196). The mean inference times per packet were 0.34633 $ms$ ($\sigma=0.2583$), 0.03849 $ms$ ($\sigma=0.3169$), 0.03704 $ms$ ($\sigma=1.018$), 0.03704 $ms$ ($\sigma=4.342$), 0.03642 $ms$ ($\sigma=20.024$), and 0.03628 $ms$ ($\sigma=28.220$) for each batch size, respectively.
The inference time for the SVM model was 0.1158 $ms$ ($\sigma=0.06$) for a single input (consisting of 5 classification outputs from SA-CNN). Thus, the total inference time for the cascaded model does not exceed 0.5 $ms$. 
Given this, we can argue that the inference time (of our models) is low enough to detect measurement traffic in a timely fashion. Also, we only need to classify the first five packets of a traffic flow to detect the measurement; all subsequent packets of the same flow can be re-routed using the IP address (of the first five packets).

In summary, our findings demonstrate that modern machine learning techniques can accurately and promptly differentiate between measurement and non-measurement packets within the 3-hop measurement circuit setup. As a result, it becomes evident that this approach cannot serve as an effective mitigation measure.

\subsection{FlashFlow}
\vspace{-0.20cm}
Traudt et al.~\cite{traudt2021flashflow} demonstrated through experiments that the true capacity of Tor network is underestimated by around 50\% due to imperfections in the currently deployed bandwidth measurement mechanism in Tor. As a better alternative, they proposed a new method called \emph{FlashFlow}, which forces relays to demonstrate their near-maximum capacity. In particular, measurements in FlashFlow are performed by teams of \emph{measurers}, each coordinated by their own authority, referred to as \emph{BWAuth}, who orchestrates the measurement process and aggregates the results. The target relay is simultaneously measured by a team of measurers over specially constructed measurement circuits, which are created and handled differently from the circuits utilized by regular client traffic. In particular, measurers and target relays are connected through single-hop connections established with the help of BWAuths and their public keys. Remarkably, each relay will only accept connections from a given BWAuth and its team once per measurement period, which implies that requests for the measurement circuit establishment can be denied for legitimate reasons. 
It is evident that identifying measurement traffic and dropping non-measurement data flows in FlashFlow is even more straightforward than in currently deployed SWBS or TorFlow. Indeed, since the measurement and non-measurement circuits are handled differently by design, they can be trivially distinguished. Consequently, the client traffic can be dropped as required by \cosys{}, or the measurement traffic can be redirected to a powerful dedicated server as in \desys{}. 
Further, FlashFlow allows an attacker to either accept or refuse being measured by specific BWAuths based on their preference. 
As a result, an attacker with several relays is able to schedule the measurements in such a way that all relays will be measured at different times. This avoids co-measurements, and further increases the inflation factor, strengthening our proposed attack.

\subsection{Other Solutions}
\label{subsec:other_solutions}
\vspace{-0.20cm}
Next, we outline some other alternate measurement solutions in the literature for Tor and discuss them in relation to our proposed \sys{} attacks. Although these solutions do not directly impact the efficacy of our proposed attacks, they are still vulnerable to inflation attacks such as \sys{}, either due to their own issues as highlighted below 
or their dependency on existing bandwidth measurement mechanisms such as TorFlow.

\noindent\textbf{EigenSpeed.}
Snader and Borisov \cite{snader2009eigenspeed} proposed a peer-to-peer (P2P) bandwidth evaluation technique for Tor in which each relay observes and stores bandwidth attained while communicating with other relays in a vector. The vector is then forwarded to an authority that uses \emph{Principal Component Analysis (PCA)} to produce consensus bandwidths for all Tor relays. Although EigenSpeed is not vulnerable to \sys{} due to the absence of specialized measurement traffic, it has been shown that EigenSpeed is vulnerable to Sybil attacks, which also leads to inflation attacks \cite{mitseva2020}.

\noindent\textbf{PeerFlow.}
This technique builds on the same P2P principle as EigenSpeed and uses two measurement techniques where each relay records the amount of traffic sent and received to/from other relays while also reporting their own available bandwidth. These measurements then get forwarded to the DAs who compute a consensus weight based on these reported values and by employing a set of trusted relays to verify them \cite{johnson2017peerflow}. PeerFlow continues to use TorFlow during bootstrapping of new Tor relays and thus attacks on TorFlow will still have a considerable impact here making PeerFlow still vulnerable to \sys{} \cite{mitseva2020}.

\noindent\textbf{MLEFlow.}
This technique employs a maximum likelihood estimation (MLE) approach to estimate the real capacity of a relay by using a series of measurements and consensus weights published by the BAs \cite{darir2022mleflow}. The authors of MLEFlow were able to achieve estimation errors below 5\% in contrast to \emph{TorFlow} and \emph{SBWS} which have estimation errors over 20\%. However, similar to the above techniques, their technique also relies on BA-based measurements, thus making it vulnerable to measurement traffic detection and inflation attacks such as \sys{}.

\section{Countermeasures}
\label{sec:countermeasures}
We now briefly outline measures that could be potentially used to mitigate \sys's bandwidth inflation strategies against Tor and discuss the challenges surrounding them.

\subsection{Detecting Measurement Anomalies}
\label{history_and_probing}
\vspace{-0.20cm}
As demonstrated by us in \cref{sub:inflation_factor}, when two or more (adversarial) relays residing on the same machine are measured simultaneously, a noticeable drop in the instantaneous bandwidth measurement (related to those relays) is observed, which is recorded and archived in the bandwidth files by the measuring BAs.  
Given that Tor archives all the bandwidth files generated since 2017, such trends indicating sudden drops in measured bandwidths of relays during co- or simultaneous measurements are easily detectable (by network BAs/DAs) by simply looking at the historical data inside the bandwidth files and can be used to flag suspicious relays. 
As mentioned in \cref{sub:inflation_factor}, bandwidth files only contain end timestamps of the measurements, making it difficult to accurately recreate the timeline of relay co-measurements.
In order to overcome this, BAs can record the start of measurement timestamp information for their individual probes and can collaborate with other BAs to successfully recreate the co-measurement timelines from the bandwidth files to identify suspicious relays that may be involved in such bandwidth inflation attacks.
BAs can then actively and collaboratively probe such suspicious relays by measuring them simultaneously in order to confirm bandwidth measurement anomalies and detect cases of bandwidth inflation. 
Moreover, as each BA is able to run at least four measurement threads in parallel, even individual BAs (without collaboration with other BAs) can perform such active probing to detect cheating or malicious relays.

\subsection{Obscuring Measurement Traffic}
\label{obsecuring_measurements}
\vspace{-0.20cm}
Another mitigation strategy that could be employed by the BAs is to obfuscate measurement traffic, say by concealing or randomizing the origin and/or destination information (of the measurement flows), with the goal of making the detection of such traffic non-trivial and difficult. Such an approach seems plausible as the detection of measurement traffic is typically the first step in most bandwidth inflation attacks, including \sys{}. 
To this end, one option is to randomize the network (IP) address of the measurement file download server. However, file downloads (during bandwidth measurement) from these servers could produce uniquely identifiable traffic flows, which in turn could be used to detect measurement traffic (by the relays). For instance, in current Tor measurement mechanisms (e.g., in TorFlow and SBWS), the download throughput is swiftly increased at the start of the measurement until it reaches a constant state to measure the maximum possible throughput of the relay, followed by rapidly reducing it towards the end of the measurement. 
As demonstrated by us in \ref{traffic_classification}, traffic classification is still possible even under masked IP addresses. We were able to identify measurement traffic from non-measurement traffic with only 5 packets (and only using their headers) from a network flow with over 0.99 precision and recall. 
Similar works in literature have also shown that traffic flow patterns can be used to achieve traffic classification using machine/deep learning mechanisms with accuracies over 95\% \cite{arestrom2019early,salman2018multi}.
These research efforts show that obfuscation or randomization of download file server IP addresses, although intuitive, may not always be an effective mitigation strategy against bandwidth inflation attacks.

\subsection{Co-residency Detection}
\label{coresidency_detection}
\vspace{-0.20cm}
An increase in traffic could lead to the relays being overloaded and such relay overloads are self-reported in Tor Metrics platform \cite{collector}. From an attack mitigation point of view, it is easy to see that such relay overload indicators can be employed to detect potential co-resident relays and neutralize bandwidth inflation attacks, including the proposed ones by us.
However, since the overload status is self-reported, one possible workaround to override this default overload reporting behavior is to employ a modified (non-standard) Tor client code which conceals the overload-related information from being reported.
Nonetheless, this form of detection is only possible as long as attackers are unaware of this detection method. 
In addition to the above, usage of the same public IP addresses and port numbers and exhibiting similar uptime and measurement results could also be indicative of co-residency.

Another possible indicator of co-residency could be relays running the same version of Tor, assuming the attacker installs these relays together or around the same time period (Tor currently has 44 different version releases in operation) \cite{tor_versions}.

\subsection{Eliminating Explicit Measurement Traffic}
\vspace{-0.2cm}
As discussed earlier, the presence of measurement traffic inherently makes the Tor network vulnerable to bandwidth inflation attacks due to the high possibility of measurement traffic detection. 
As a result, perhaps one of the most promising countermeasures is to eliminate the use of explicit measurement traffic to accomplish bandwidth measurement.  To this end, one approach would be to use the existing Tor relay nodes themselves to measure their peers' bandwidth in a distributed fashion by using regular Tor client traffic.
It must be noted that such a technique has already been proposed by some of the existing TorFlow/SBWS alternatives proposed in the literature, such as EigenSpeed \cite{snader2009eigenspeed} and PeerFlow \cite{johnson2017peerflow} (as discussed earlier in \cref{subsec:other_solutions}), which employ a distributed (or peer-to-peer) network of measurements nodes. 
Although these solutions are currently incompatible with Tor and have their own shortcomings, as discussed in \cref{sec:related_works} and \cref{sec:res_alt_solutions}, we believe that improved versions of these solutions could be effective in hindering a wide range of bandwidth inflation attacks including \sys{}.

\section{Related Works} 
\label{sec:related_works}
Bandwidth inflation in Tor was initially exploited back in 2006 by Overlier et al. \cite{overlier2006locating}. At the time, relays reported their own bandwidth measurements to DAs, and the authors reported inflated measurements to increase the weight of malicious relays in an attempt to achieve more efficient attacks against Tor hidden services. In 2007, Bauer et al. \cite{bauer2007low} exploited the same design flaw to propose a traffic analysis attack that affects client anonymity and showed that a low-resource adversarial entity could falsely report inflated bandwidths allowing them to acquire both entry and exit relays of the same client with an increased probability. 

Consequently, bandwidth authorities and scanning protocols (TorFlow) were introduced in Tor in 2011. TorFlow was gradually replaced by SBWS starting 2018.
However, in 2013 Biryukov et al.  \cite{biryukovTrawling2013} showed that an inflation attack is still feasible by detecting bandwidth measurement traffic and prioritizing the entire available bandwidth to such traffic while throttling bandwidth for all other traffic. They were able to achieve more than tenfold (10x) bandwidth inflation using this strategy.
Johnson et al. \cite{johnson2017} also confirmed this attack in a simulated environment and were able to achieve an 177$\times$ inflation compared to the actual bandwidth where the alteration achieved for the consensus weight increased from 7\% to 11\%. They further showed that the bandwidth utilized by the adversarial relay dropped from 22.5 \emph{MBps} to a mere 0.2 \emph{MBps} which also validates that an attacker does not require heavy resources to perform such an attack. 
They further exposed a vulnerability in EigenSpeed which allows an attacker to kick out honest relays by adding a very large number of malicious relays (or Sybils). 
They used the same malicious relays (or Sybil nodes) to report false bandwidths for honest relays and obtained a 28$\times$ bandwidth inflation.

More recently, Mitseva et al. \cite{mitseva2020} conducted a practical analysis of PeerFlow \cite{johnson2017peerflow}.
They observed that since PeerFlow uses TorFlow measurements in their initial phase when new relays join the Tor network, it is vulnerable to inflation attacks up to 9$\times$ times compared to the specified security boundary of 4.6$\times$ in PeerFlow.
Mitseva et al. also identified a design flaw in PeerFlow due to its variable measurement period times, leading to malicious new relays compelling to be measured using TorFlow by avoiding connections to other relays. 

All of these studies share similarities with our work, as they also delve into various strategies for bandwidth inflation, aiming to maximize user traffic routed through malicious relays while minimizing resource usage. Consequently, these strategies facilitate a wide range of privacy attacks in Tor, including website fingerprinting \cite{abe2016fingerprinting, rimmer2017automated, jansen2018inside, cherubin2022online}, flow correlation \cite{houmansadr2009rainbow, yu2007dsss, wang2007network, nasr2018deepcorr}, and routing attacks \cite{bauer2007low, sun2015raptor}. However, our approach distinguishes itself from previous methods by introducing a novel inflation attack technique. By combining this technique with existing approaches, attackers can achieve even more significant bandwidth inflation.

\section{Conclusion}
\label{sec:conclusion}
In this work, we presented a novel inflation attack on Tor which adversarial relays could use to inflate their bandwidth. 
We evaluated two attack variants, a co-residing relay attack (\cosys{}) which achieved an inflation gain of nearly $n$, and a dedicated server attack (\desys{}) which achieved an inflation multiplier by nearly half $n*N$, where $n$ is the size of the relay cluster and $N$ the number of servers.
Next, we explored the coincidence rate, which will help an attacker to optimally choose the number of co-resident relays to maximize their inflation factor. 
We found that the inflation factor increases almost linearly for cluster sizes up to 120 relays.
In our theoretical examination, we demonstrate that by utilizing only 10 specialized servers employing the \sys{} attack, each having a bandwidth of 100MB/s, it is possible to gain command over 50\% of the Tor network's traffic.
We also provided further insights on Tor relating to relay families and port usage distribution. 
Finally, we also demonstrated how \sys{} is resilient against other alternative measurement techniques, followed by potential countermeasures against the proposed bandwidth inflation attacks.

\bibliographystyle{plain}
\bibliography{references}

\appendices
\label{sec:appendix}

\section{Observed Relay Families}
\label{sec:app:relay-families}

The relay families (located in Germany) we identified during preliminary investigations with potentially co-residing relays are shown in \cref{tab:families}.
\begin{table}[h]
\centering
\caption{A List of Observed Relay Families.}
\label{tab:families}
\begin{adjustbox}{width=0.98\linewidth}
\begin{tabular}{ccc}
\toprule
Family & No. of Relays & Bandwidth (Mbps) \\ 
\midrule
8029928877A15A504D92996C65BFD4F0BDF4E702 & 88 & 2596.32 \\
2DF03D7B158DAE2EAF76078775451F1769506451 & 85 & 2003.7 \\
5B83DC983406651A0B4F6AE1940793CDD6A6F92E & 56 & 1784.92 \\
526AD50C9DE6AF533DEBE8F9BBDF149BC1F5AB6E & 178 & 1278.92 \\
A4E47F08B8D56428DF76B17EDD6738BCBC3F5EFB & 55 & 1092.18 \\
135F2A8B32F583845F2B0E133EFD84C25026761B & 32 & 1077.46 \\
1050FC79C5F1103B185300EF72DDF5B4EDC683C9 & 94 & 936.37 \\
\bottomrule
\end{tabular}
\end{adjustbox}
\end{table}

\section{2-hop vs. 3-hop Measurement Circuits}
\label{sec:app:2vs3hop}

\begin{figure}[h]
\includegraphics[width=\linewidth]{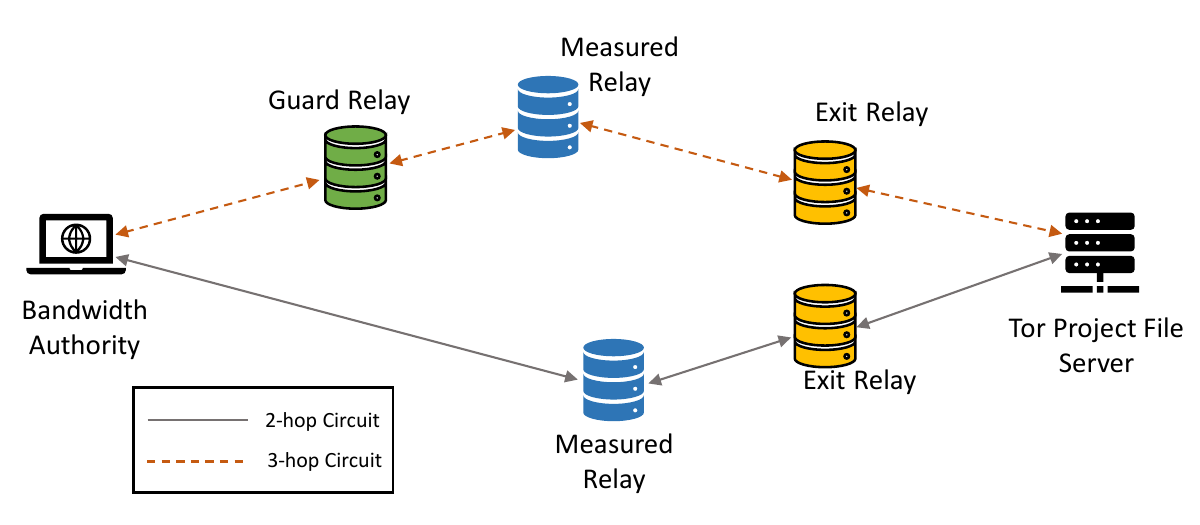}
\centering
\vspace*{-0.375cm}
\caption{2-hop vs. 3-hop measurement circuits}
\label{fig:3hop}
\vspace*{-0.375cm}
\end{figure}

The difference between a 2-hop and a 3-hop measurement circuit can be seen in \cref{fig:3hop} as previously described in \cref{traffic_classification}.

\section{Attack Resource Estimation Equations}
\label{sec:app:resource_forumlae}

As mentioned in \cref{sub:opt_resources}, the curve fitting algorithm to construct a matching curve from the inflation factor of the attack is defined in \cref{eq:inflation_factor}. 
The fitted function (mean squared error is 0.0318 and coefficient of determination is 0.99995) returns the inflation factor for a cluster size, $x$. 

The equation to calculate the required amount of dedicated servers in the \desys{} attack, given a cluster size and how much traffic should be controlled, is shown by the equation \cref{eq:server_amount}.
The parameter $x$ refers to the cluster size, $b$ is a constant that defines the total traffic in the Tor network, $p$ defines the desired percentage of controlled traffic, and $d$ defines the bandwidth of the dedicated servers.
To calculate the perfect cluster size, one can create a linear program (LP) that minimizes both the cluster size and the number of dedicated servers (i.e., $\mathrm{minimize}\ x + s(x,b,p,d)$).

\begin{gather}\label{eq:inflation_factor}
i(x) = 0.75895138 \cdot (1.44995314 \cdot x)^{0.96837148}\\\nonumber
- (0.03714758 \cdot x)^{2} - 0.07672455\\\nonumber
\mathbf{D}_{i} = \mathbf{N} \in [1, 120]
\end{gather}

\vspace*{-0.375cm}

\begin{gather}\label{eq:server_amount}
s(x,b,p,d) = \Big\lceil\frac{2 \cdot b \cdot \frac{p}{100}}{d \cdot i(x)}\Big\rceil\\\nonumber
\mathbf{D}_{s} = \mathbf{N}^4,x \in [1, 120],p \in [1, 100]
\end{gather}

\section{Distribution of Ports by Tor Relays}

\begin{figure}[H]
\includegraphics[width=0.7\linewidth]{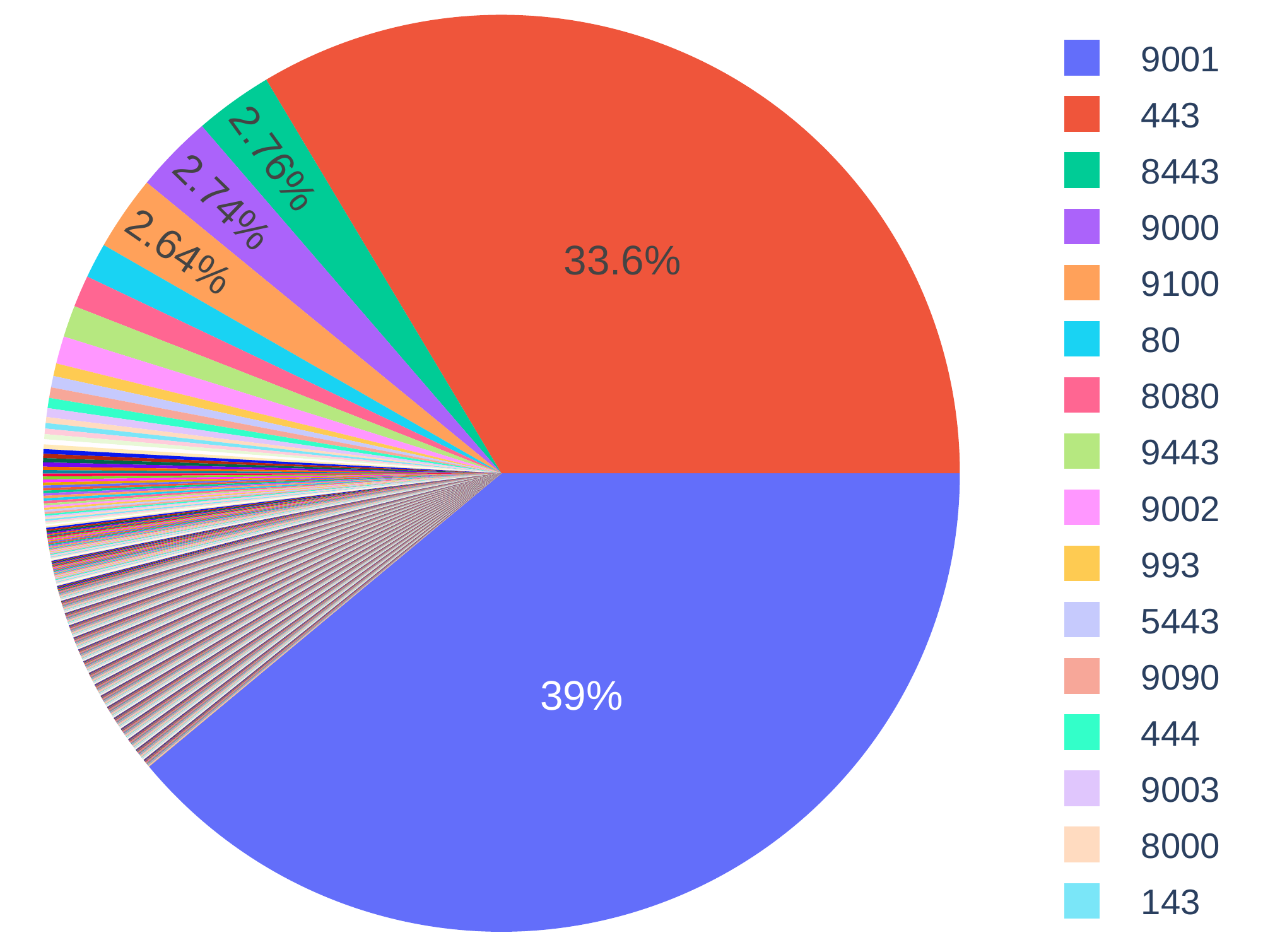}
\centering
\caption{Distribution of different ports used by Tor relays.}
\label{fig:port_usage}
\vspace*{-0.375cm}
\end{figure}

The analysis done on the real Tor network shows 581 different port numbers being used by relays, suggesting possible co-resident relays (\cref{fig:port_usage}).

\end{document}